\documentclass[a4paper,11pt]{article}
\pdfoutput=1 

\usepackage{jheppub} 

\usepackage[utf8]{inputenc} 
\usepackage[T1]{fontenc} 

\usepackage{color}
\usepackage{amsmath}
\usepackage{comment}
\usepackage{mathrsfs}

\font\mybb=msbm10 at 11pt

\font\mybbsub=msbm10 at 8pt

\def\bb#1{\hbox{\mybb#1}}

\def\bbsub#1{\hbox{\mybbsub#1}}

\def\FF {\bb{F}}
\def\FFsub{\bbsub{F}}

\def\ZZ {\bb{Z}}

\def\PP {\bb{P}}
\def\PPsub{\bbsub{P}}

\def\Fn{\FF_n}

\newcommand\beqa{\begin{eqnarray}}
\newcommand\eeqa{\end{eqnarray}}
\newcommand\n{\nonumber\\}

\def\beq#1\eeq{\begin{equation}#1\end{equation}}
\def\bes#1\ees{\begin{equation}\begin{split}#1
               \end{split}\end{equation}}
\def\bea#1\eea{\begin{align}#1\end{align}}

\begin{document}

{~}
\title{Matter from multiply enhanced singularities in F-theory}

\author[a]{Shun'ya Mizoguchi}
\author[b]{and Taro Tani}

\affiliation[a]{Theory Center, Institute of Particle and Nuclear Studies, KEK\\
Tsukuba, Ibaraki, 305-0801, Japan}
\affiliation[b]{National Institute of Technology, Kurume College, \\Kurume, Fukuoka, 830-8555, Japan}
\emailAdd{mizoguch@post.kek.jp}
\emailAdd{tani@kurume-nct.ac.jp}

\abstract{

We investigate the geometrical structure of multiply enhanced codimension-two singularities 
in the $SU(5)$ model of six-dimensional F-theory, where the rank of the singularity increases 
by two or more. We perform blow-up processes for the enhancement $SU(5)\rightarrow G'$, 
where $G'=E_6$, $E_7$ or $E_8$, to examine whether a sufficient set of exceptional curves 
emerge that can explain the charged matter generation  predicted from anomaly cancellation. 
We first apply one of the six Esole-Yau small resolutions to the multiply enhanced singularities, 
but it turns out that the proper transform of the threefold equation does not reflect changes 
in the singularity or how the generic codimension-two singularities gather there.
We then use a(n) (apparently) different way of small resolutions than Esole-Yau to find that,
except for the cases of $G'= E_6$ and special cases of $E_7$,  either (1) the resolution 
 only yields exceptional curves that are insufficient to cancel the anomaly, or 
 (2) there arises a type of singularity that is neither a conifold nor a generalized conifold singularity.
Finally, we revisit the Esole-Yau small resolution and show that the change of the way 
of small resolutions amounts to simply exchanging the proper transform and the constraint condition, 
and under this exchange the two ways of small  resolutions are completely equivalent.

}

\preprint{KEK-TH-2637}
\date{July 20, 2024 (Dec 23, 2024)}
\maketitle

\section{Introduction}

There is no doubt that singularities play an essential role in F-theory.
It is basically formulated on an elliptically fibered complex manifold on some base, 
where the elliptic fiber modulus represents the value of the complex scalar field $\tau$ 
of type IIB string theory \cite{Vafa}.  Depending on its complex structure, 
a singularity may arise 
somewhere on the fiber. From the perspective of type IIB theory compactified on the base, 
this is where a 7-brane exists. More precisely, a 7-brane resides at a locus on
the base over which the fiber develops a singularity.  As branes overlap or intersect 
each other to achieve gauge symmetry or matter in string theory \cite{MV1,MV2,BIKMSV},
the structure of the singularities in F-theory determines how they are 
geometrically realized.

On an elliptic {\em surface}, 
the types of singularities 
are classified by Kodaira \cite{Kodaira}. Precisely speaking, 
it classifies the types of the set of 
exceptional fibers that arise through the process of resolving a singularity.
As is well known, 
Kodaira's fiber type is specified by, except for a few exceptional cases, 
an extended Dynkin diagram of some simple Lie algebra representing 
how $\PP^1$'s arising as exceptional sets intersect each other.
So each fiber type is naturally associated with some 
simple Lie algebra $G$; its fiber type is often specified 
by this Lie algebra $G$ or its corresponding  Dynkin diagram, just like  
the du Val-Kleinian $ADE$ singularities, rather than by the original name 
of Kodaira's fiber type.
We will also use this terminology to refer to these singularities 
in this paper.
In the literature, such a singularity is 
often called a codimension-one singularity\footnote{\label{footnote1}
We should note that the usage of this term
is somewhat misleading; what is actually codimension-one 
is the locus on the {\em base} of the elliptic 
fibration, but the actual singularity lies on the {\em fiber} over such a locus.
Therefore, while this terminology is appropriate for the type IIB 
(or F-GUT \cite{DonagiWijnholt,BHV1,BHV2,DonagiWijnholt2}) perspective,
from the F-theory perspective it would be more reasonable to 
say that this singularity is {\em codimension-two}, counting the 
codimensions in the total elliptic space.
However, in this paper, following \cite{EsoleYau}, 
we will (somewhat misleadingly) refer to this as a {\em codimension-one} singularity, 
which is the same terminology as in much of the literature on F-GUT.
}.

If one fibers such elliptic surfaces over some base manifold $\cal B$, 
one can obtain a higher-dimensional elliptic manifold whose base is 
a $\PP^1$ fibration over ${\cal B}$, on which one can consider F-theory.
In six-dimensional F-theory, the base of the elliptic fibration 
is complex two-dimensional, so 
codimension-one discriminant loci 
intersect each other 
generically at isolated points. 
As is well known, matter emerges 
from these  codimension-two singularities\footnote{Again, what is codimension-two is the locus on the base, and the singularity itself on the elliptic manifold is codimension three. Still, we call it {\em codimension-two} similarly.} 
\cite{KatzVafa,BIKMSV}, and on the fiber over such loci, the $G$ singularity over the generic codimension-one loci changes into a different $G'$ singularity of higher rank.
The change in the rank from $G$ to $G'$ is one generically.


In this paper, we consider codimension-two singularities 
in six-dimensional F-theory, 
where the rank of the singularity $G'$ is two or more higher 
than that of the generic co-dimension-one singularity $G$ around it. 
We will call such a change in the rank of the 
Lie algebras of the singularities ``{\em multiple enhancement}''. 
In six dimensions, these multiply enhanced 
codimension-two singularities do not generically occur, 
but appear only when the complex structure moduli are specifically tuned
so that several generic codimension-two singularities happen to overlap.
In the intersecting 7-brane picture, this corresponds to the case 
where two branes no longer intersect transversally.

In a generic rank-one enhancement from $G$ to $G'$ 
in six dimensions, 
massless matter arising there can be understood as coming from light 
M2-branes \cite{KatzVafa} wrapped around extra two-cycles that newly emerge
over the enhanced point. 
An understanding of matter generation in terms of string junction was 
discussed in \cite{Tani}.
In typical cases, these massless states 
correspond to one of the roots of $G'$. In particular, 
 if there is no monodromy among these two-cycles 
(the ``split'' case), the massless matter arising there is given by $G'/(G\times U(1))$,
which forms a homogeneous K\"ahler manifold. 
As was investigated in detail in \cite{BIKMSV}, if the base space of 
the elliptic fibration is taken to be a Hirzebruch surface \cite{MV1,MV2}, 
the set of matter multiplets given by this rule coincides exactly with 
those of $E_8\times E_8$  heterotic string 
on a K3 surface, in particular all anomalies cancel. 
On the other hand, 
as mentioned above, a singularity that enhances 
the rank by two or more occurs when several ordinary rank-one enhanced singularities 
come together and overlap. This means that, at such special points on the 
moduli space, the massless matter hypermultiplets arising from some of the 
ordinary rank-one enhanced singularities that existed disappear, and a new set of 
massless matter hypermultiplets is created at the multiply enhanced 
singularity. 

In six dimensions, however, the constraints for 
anomaly cancellation are very strict so that the changes in the spectrum of 
massless matter are severely restricted. Indeed,  
anomaly cancellation requires 
that there should be no net change in the number of hypermultiplets in such transitions \cite{MTanomaly}. 
This is simply because the Green-Schwarz mechanism in six dimensions \cite{GSW,Sadov}
requires that the numbers of vector, tensor and hypermultiplets 
$n_V$, $n_T$ and $n_H$ satisfy 
\beqa
n_H- n_V&=&273 - 29 n_T.
\eeqa
Therefore, the number of hypermultiplets cannot change unless the number of multiplets in vectors and tensors does not change (which is assumed here)\footnote{In general,
a different representation contributes differently to the anomalies, so the number 
of hypermultiplets in each representation must be the same before and after the transition.}.
This prohibition on the change in the number of hypermultiplets 
in turn predicts what charged hypermultiplets occur from 
such a singularity with a multiple enhancement. 
The purpose of this paper is to examine whether the massless 
matter predicted in this way can be understood from the geometric structure 
of the singularity with a multiple enhancement.

In this paper we revisit the well-studied compactification of six-dimensional F-theory on Hirzebruch surfaces $\FF_n$ \cite{MV1,MV2,BIKMSV} with unbroken $SU(5)$ gauge symmetry.
We will explicitly carry out the resolution process in the cases of multiple enhancements 
$SU(5)\rightarrow E_6$, $E_7$ and $E_8$ to see whether the expected 
matter spectrum can be explained in terms of exceptional curves 
arising from the resolutions.
We are interested in the local structure of the singularity, so although we 
work in this particular global set-up, the result will apply to any six-dimensional 
F-theory compactification.

The singularity structure of the {\em four}-dimensional F-theory $SU(5)$ model compactified 
on a Calabi-Yau fourfold was analyzed in \cite{EsoleYau}, and very impressive results were revealed.
(See also \cite{relatedanalysis1,relatedanalysis2,relatedanalysis3,relatedanalysis4,relatedanalysis5}.)
Starting with the $SU(5)$ equation in Tate's form, two-time codimension-one blow-ups yield
an equation in the ``binomial'' form, that is,
the equation for the threefold after the two blow-ups takes the form
\beqa
u_1u_2u_3-v_1v_2&=&0, \label{v1v2-u1u2u3=0}
\eeqa
where $u_1,\ldots,v_2$ are sections of some relevant line bundles 
(see later sections for details).
The equation (\ref{v1v2-u1u2u3=0})
indicates that, if $v_1$, $v_2=0$ and two (and only two) of the three sections  
$u_1$, $u_2$ and $u_3$ simultaneously vanish, there is a conifold singularity.
This is the place where a codimension-two singularity resides;
after the codimension-one blow-ups, conifold singularities remain at the codimension-two 
loci (in the sense of the base as we remarked). Note that, in four-dimensional F-theory 
considered in \cite{EsoleYau}, these conifold singularities form complex curves 
such as discussed in \cite{{ConifoldTransitionsinMtheory}}.

It was found in \cite{EsoleYau} 
that, generically, these 
conifold singularities are all disingularized by successive two small resolutions 
inserting two curves of $\PP^1$'s replacing the curves of singularities. 
There are six ways to do this, depending on which of $u_i$'s  is paired with $v_1$
 or $v_2$. For instance, this is done by replacing, say, $(v_1,u_1)$ and 
  $(v_2,u_2)$ with 
\beqa
v_1=\xi V_1,~~~u_1=\xi U_1,
\label{EYsmallresolution1}\\
v_2=\zeta V_2,~~~u_2=\zeta U_2,
\label{EYsmallresolution2}
\eeqa
where 
$(V_1:U_1)$ and $(V_2:U_2)$ are the homogeneous coordinates of $\PP^1$'s 
mentioned above, and $\xi$ and $\zeta$ are sections of appropriate 
line bundles served for projectivization\footnote{$\xi$, $\zeta$ here 
were denoted as $\delta_3$, $\delta_4$ in \cite{Yukawas}, respectively.}.

This also applies to the present six-dimensional case. Again, in the generic 
case where there is no multiple enhancement, all the codimension-two 
conifold singularities are resolved in this way.
However,
in the case of multiple enhancement, 
a different type of singularity appears than a conifold singularity, at which 
all three $u_i$'s vanish simultaneously. 
In fact, such singularities are also 
known to appear in generic four-dimensional models: The codimension-three $E_6$ 
singularity. In this four-dimensional case, even such non-conifold 
singularities are readily resolved by the above two small resolutions \cite{EsoleYau,Yukawas}. 
In six dimensions, on the other hand, there is no such thing as ``codimension-three''
singularity (since the base is two-dimensional) but such singularities
appear only when the complex structure is properly tuned.

We will perform the resolution of this kind of singularity in two different ways.
A difference occurs after the two codimension-one blow-ups\footnote{ 
In fact, the first small resolutions ((\ref{EYsmallresolution1}) and (\ref{Tanismallresolution1}))
performed after the codimension-one blow-ups 
are the same, so the difference arises in the second small resolutions.}.
One way is through two small resolutions (\ref{EYsmallresolution1})(\ref{EYsmallresolution2}) 
done in \cite{EsoleYau} as described above.
As we will see
, it turns out that if we do this, we run into something strange: 
Even in the case of multiple enhancements, it ends up looking like nothing special is happening. 
So we consider an alternative way of small resolutions: We perform the same 
first small resolution (\ref{EYsmallresolution1}) as above,
but for the second small resolution 
we insert $\PP^1$ at $V_1=u_3=0$:
\beqa
v_1=\xi V_1,~~~u_1=\xi U_1,
\label{Tanismallresolution1}\\
V_1=\eta V'_1,~~~u_3=\eta U_3.
\label{Tanismallresolution2}
\eeqa
%
This change of the center of the blow up leads to an equivalent small resolution 
for ordinary conifold singularities. 
However, we will see that there is indeed a difference between (\ref{EYsmallresolution2}) 
and (\ref{Tanismallresolution2}) when we actually apply the resolution procedure to 
a multiply enhanced singularity. 

We will then revisit the first way of small resolutions (\ref{EYsmallresolution2}). 
We will show what was overlooked and why it looked like ``nothing happened'' 
then. 
In fact, if we properly consider what we missed, 
we will see that the two ways of small resolutions are again equivalent for multiply enhanced singularities as well, and exactly the same conclusions can be drawn from both.
This equivalence is realized in a rather interesting way as a certain ``duality'', in which 
the proper transform of the threefold equation in one way of small resolutions 
corresponds to the constraint equation in the other.
   
%

The organization of this paper is as follows.
In section \ref{sec:mes}, we list possible realizations of singularities to
achieve each multiple enhancement 
and examine how many hypermultiplets are expected to arise there to cancel the anomalies. 
It is shown that a multiple enhancement $SU(5)\rightarrow E_n$ ($n=6,\,7$ or $8$) is realized at
a locus where some of the rank-one enhanced singularities $SU(5)\rightarrow SO(10)$ and 
$SU(5)\rightarrow SU(6)$ overlap.
For given $n$, various ways of overlapping are allowed, and 
depending on it, the anomaly-free spectrum expected to arise there varies.
If the expected spectrum is given by the homogeneous 
K\"ahler manifold $E_n/(SU(5)\times U(1)^{n-4})$, we call such a singularity a complete singularity,
whereas, if it is smaller than that, we call such a singularity an incomplete singularity.
The complete/incomplete singularities and the expected anomaly-free spectra are listed for each $E_n$
in section \ref{section:incompleteE6E7E8}.
%
In section \ref{sec:EY}, we apply the Esole-Yau small resolution (\ref{EYsmallresolution1})(\ref{EYsmallresolution2})
to those singularities and find that, in all cases,
the proper transform after the two-time codimension-one blow-ups is regular. 
In section \ref{Alternativesmallresolution}, we use the alternative small resolution (\ref{Tanismallresolution1})(\ref{Tanismallresolution2}) and show that, in this procedure, the proper transforms reflect  
changes in the singularity.
We find that, except for the case of $G'=E_6$ and special cases of $G'=E_7$, either
(1) the resolution only yields exceptional curves that are insufficient to cancel the anomaly, or 
(2) there arises a type of singularity that is neither a conifold nor a generalized conifold singularity\footnote
{One can show that these phenomena are not specific to $SU(5) \rightarrow E_n$,
but common to multiple enhancements.
In particular, a simple example of (1) is $SU(2) \rightarrow SO(8)$.}.
In section \ref{duality}, we revisit the Esole-Yau small resolution and
find that there is a ``duality'' between the two ways of small resolution, showing that 
they are equivalent.

\section{Multiply enhanced singularities in 6D F-theory}
\label{sec:mes}

\subsection{Tate's form and multiply enhanced singularities}

We consider six-dimensional compactifications of F-theory on Calabi-Yau threefolds (CY$_3$), 
which are elliptically fibered over a Hirzebruch surface $\Fn$.
Let $z$ and $w$ are affine coordinates of the fiber 
$\PP^1$ and the base $\PP^1$ in the relevant coordinate patch of $\Fn$, respectively.
We describe such a CY$_3$ in Tate's form 
\beq
   y^2 + a_1 x y +a_3 y =  x^3 + a_2 x^2 + a_4 x +a_6.
\label{eq:Tate}
\eeq
$a_i$'s are polynomials of the coordinates $(z,w)$ of $\Fn$ of particular degrees, 
representing the sections of appropriate line bundles that satisfy the Calabi-Yau 
condition.
Such a CY$_3$ can also be seen as a K3 fibered geometry over the base $\PP^1$ of $\Fn$.

To achieve a split $SU(5)$ model \cite{BIKMSV}, 
we assume the orders of the sections in $z$ as
\beq
   {\rm o}(a_1\, ,a_2,\, a_3,\, a_4,\, a_6) = (0,1,2,3,5),
\eeq
where we write the order of the polynomial $a_i(z)$ in $z$ as ${\rm o}(a_i)$.
They are realized by assuming 
\bes
 a_1 & = a_{1,0} + a_{1,1} z +\cdots,  \\
 a_2 & = a_{2,1} z + a_{2,2} z^2 +\cdots,  \\
 a_3 & = a_{3,2} z^2 + a_{3,3} z^3 +\cdots,   \\
 a_4 & = a_{4,3} z^3 + a_{4,4} z^4 +\cdots,   \\
 a_6 &=  a_{6,5} z^5 + a_{6,6} z^6+\cdots,  
\ees
where $a_{i,j}$ are polynomials of $w$.
The leading terms are sufficient to describe the relevant structure of the singularity.
The independent polynomials defining the $SU(5)$ singularity are 
\beqa
  a_{1,0}, \, a_{2,1}, \, a_{3,2}, \, a_{4,3}, \, a_{6,5},
\label{eq:SU5polynomials}
\eeqa
and (\ref{eq:Tate}) reads (to leading order in $z$ for each $a_{i}$)
\beqa
      y^2 + a_{1,0} x y +a_{3,2} z^2 y =  x^3 + a_{2,1} z x^2 + a_{4,3}z^3 x +a_{6,5} z^5.
\label{eq:Tate0}
\eeqa
The corresponding Weierstrass equation is given by 
\beqa
 y^2 = x^3 + f x + g
\eeqa
with
%
\bes
     f & = -\frac{1}{48} a_{1,0}^4 +\left(-\frac{1}{6} a_{1,0}^2 a_{2,1} +\cdots\right)z
     +\left(-\frac{1}{6}(2a_{2,1}^2-3a_{1,0}a_{3,2})+\cdots \right)z^2 \\&
     +(a_{4,3}+\cdots)z^3+\cdots, \\
    g & = \frac{1}{864} a_{1,0}^6
    +\left(\frac{1}{72} a_{1,0}^4 a_{2,1} +\cdots\right)z 
    +\left(\frac{1}{72} a_{1,0}^2(4a_{2,1}^2-3a_{1,0}a_{3,2})+\cdots\right)z^2   \\
       &   + \left(\frac{1}{108}(8a_{2,1}^3-18a_{1,0}a_{2,1}a_{3,2}-9a_{1,0}^2 a_{4,3})+\cdots\right)z^3\\ 
         & + \left( \frac{1}{12}(3a_{3,2}^2-4a_{2,1}a_{4,3})+\cdots\right)z^4 +(a_{6,5} +\cdots)z^5+\cdots.
\label{eq:Weierstrass}
\ees
The discriminant is given by 
\bes
 \Delta & = 4f^3+27g^2 \\
            & = \frac{1}{16} \big( a_{1,0}^4 \, P_{8,5} \, z^5 + a_{1,0}^2 \, Q_{10,6} \, z^6 + R_{12,7} z^7 + \cdots  \big),
\label{eq:Discriminant}
\ees
where 
\bes
    P_{8,5} & = a_{2,1}a_{3,2}^2-a_{1,0}a_{3,2}a_{4,3}+a_{1,0}^2 a_{6,5},  \\
    Q_{10,6} & =  8 a_{2,1}^2 a_{3,2}^2 + (\mbox{terms divisible by} \,\, a_{1,0}), \\ 
    R_{12,7} &= 16 a_{2,1}^3 a_{3,2}^2 +  (\mbox{terms divisible by} \,\, a_{1,0}).
\label{eq:PQR}
\ees
The orders of $f, g$ and $\Delta$ are generically given by $\mbox{o}(f,g,\Delta) = (0,0,5)$ 
and the $SU(5)$ singularity is
realized at the codimansion-one locus $z=0$.
%
%

The degrees in $w$ of the independent polynomials \eqref{eq:SU5polynomials} 
depend on $n$ of $\Fn$ and are given by 
\beq
    \mbox{deg}(a_{i,j}) = n\,(i-j)+2i.
\label{eq:aijorders0}
\eeq
%
Writing explicitly:
\beq
 \mbox{deg}(a_{1,0},\,a_{2,1},\,a_{3,2},\,a_{4,3},\,a_{6,5}) = (n+2,\, n+4,\, n+6,\, n+8,\, n+12).
\label{eq:aijorders}
\eeq
The number of neutral hypermultiplets is thus
\beq
  n_{H0} = (n+3)+(n+5)+(n+7)+(n+9)+(n+13)-1 = 5n + 36.
\label{eq:SU5H0}
\eeq

Charged hypermultiplets are localized at codimension-two discriminant loci, where the singularity gets enhanced. 
At $a_{1,0}=0$, one can see from \eqref{eq:Weierstrass} and \eqref{eq:Discriminant} 
that $\mbox{o}(f,g,\Delta) = (2,3,7)$, thus the enhancement $SU(5) \rightarrow SO(10)$ occurs
(see Table \ref{singorder} for the Weierstrass orders in the Kodaira classification).
Correspondingly, a chiral matter in $\bold{10}$ is localized at each point of $a_{1,0}=0$.
Also, at $P_{8,5}=0$, $\mbox{o}(f,g,\Delta) = (0,0,6)$  and the enhancement
$SU(5)\rightarrow SU(6)$ occurs. It gives a chiral matter in $\bold{5}$ localized at $P_{8,5}=0$. 
Note that both of them are rank-one enhancements.
Since $a_{1,0}$ and $P_{8,5}$ have degrees $n+2$ and $3n+16$, respectively, we obtain the charged matter spectrum
\beq
    (n+2) \bold{10} \oplus (3n+16) \bold{5}.
\label{eq:SU5Hc}
\eeq
The number of the charged hypermultiplets is thus
\beq
  n_{Hc} = (n+2)\times 10 +  (3n+16)\times 5  = 25n+100,
\eeq
which gives the number of hypermultiplets as
\beq
 n_{H} = n_{H0} + n_{Hc} = 30n+136,
\eeq
satisfying the anomaly-free condition \cite{MV1}
\beq
     n_{H}-n_{V} = 30n+112
\eeq
with the number of vector multiplets $n_V = 24$ for $SU(5)$. 

Additional conditons on $a_{i, j}$'s yield multiple enhancements (rank $\geq$ 2 enhancements). 
The conditions for realizing the enhancements to $SO(12)$ and $E_n$ ($n=6,7,8$) can be read from  \eqref{eq:Weierstrass}, \eqref{eq:Discriminant} and \eqref{eq:PQR} and are summarized in Table \ref{singorder}.
\begin{table}[tbp]
\centering
\begin{tabular}{c||ccc||l}
Singularity       &  $\mbox{o}(f)$ & $\mbox{o(g)}$ & $\mbox{o}(\Delta)$ & Conditions for $a_{i,j}$ (Tate's orders) \\  \hline \hline      
$SU(5)$    &     0    &    0     &    5    &          \\ \hline
$SU(6)$    &     0    &    0     &    6    &    $P_{8,5}=0$      \\  
$SO(10)$  &     2    &    3     &    7    &    $a_{1,0}=0$       \\    \hline  
$SO(12)$  &     2    &    3     &    8    &    $a_{1,0} =a_{3,2}=0$       \\  
$E_6$       &  $\geq$ 3    &    4     &    8    &     $a_{1,0} =a_{2,1}=0$      \\
$E_7$       &     3    &    $\geq$ 5     &    9    &   $a_{1,0} = a_{2,1}=a_{3,2} = 0$       \\
$E_8$       &   $\geq$ 4    &    5     &    10    &   $a_{1,0} = a_{2,1}=a_{3,2} = a_{4,3} = 0$    
\end{tabular}
\caption{rank-one and rank$\geq 2$ enhancements of $SU(5)$.}
\label{singorder}
\end{table}

\subsection{Incomplete/Complete multiply enhanced singularities}
Even if the Lie algebra to which the singularity is enhanced is specified, there are variety of possibilities in achieving the enhancement in general. 
For example, let us consider the rank-two enhancement $SU(5)\rightarrow E_6$. 
This enhancement is a generic one in four dimensions, but in six dimensions it 
only occurs if the complex structure is so tuned, as we mentioned. 
The condition for this enhancement is
\beqa
a_{1,0} =a_{2,1}=0.
\eeqa
Since $a_{1,0}=0$ is the condition for the enhancement to $SO(10)$, 
this is where a {\bf 10} hypermultiplet arises. If $a_{2,1}=0$
is further satisfied, $P_{8,5}$ will also become $0$ (see \eqref{eq:PQR}), so this is also 
the place where a {\bf 5} appears as localized matter. Thus we see that 
an $E_6$ point\footnote{We say that a point on the two-dimensional base 
($=\FFsub_n$ in the present case) of the elliptic fibration is a {\it $G$ point}
if the elliptic fiber over that point develops a $G$ singularity 
in the sense of Kodaira as a singularity of a surface (in the present case 
the fiber K3 surface of the K3 fibration over the base $\PPsub^1$ of $\FFsub_n$).}
can be made up of an $SO(10)$ point and an $SU(6)$ point overlapping each other.
Since the condition only requires that $a_{1,0}$ or $a_{2,1}$ 
be zero there, their orders in $w$ are arbitrary. 
Therefore, we may alternatively assume that $a_{1,0}$ has a double root 
${\rm o}_w(a_{1,0})=2$ there, 
where we denote the order of the polynomial $a_{i,j}(w)$ by ${\rm o}_w(a_{i,j})$.
Then, a slight deformation of the complex structure 
will result in two single roots of $a_{1,0}=0$ that are close to each other, at each of which 
a hypermultiplet {\bf 10} occurs. Put in the reverse direction, such a multiply enhanced 
point arises from two $SO(10)$ points and one $SU(6)$ point. Therefore, two  {\bf 10}
and one {\bf 5} must be generated there to remain free of anomalies.

This is reminiscent of the case of the ordinary rank-one enhancement 
from $SU(6)$ to $E_6$, where a generic codimension-two
singularity generates half-hypermultiplets \cite{MT}. 
In that case, let $t_r$ be the relevant section (in the notation of \cite{halfhyper}), 
then if ${\rm o}_w(t_r)=1$, a single half-hypermultiplet {\bf 20} of $SU(6)$ appears,
while  if ${\rm o}_w(t_r)=2$, there arise two half-hypermultiplets to form a full 
hypermultiplet.
It was found \cite{MT} that, in the latter case, there appears an additional conifold singularity, 
and an extra exceptional curve arising through the resolution ``completes'' the full 
$E_6$ Dynkin diagram.
A related analysis was done in \cite{Yukawas},
and other cases where massless half-hypermultiplets are generated were investigated 
in \cite{halfhyper}.

In the present $SU(5)$ models, if the enough number of exceptional curves 
arise to form the full $E_6$ Dynkin diagram, then one might similarly expect that  
\beqa
E_6/(SU(5)\times U(1)^2)&=&{\bf 10}\oplus{\bf 10}\oplus{\bf 5}\oplus{\bf 1}
\label{eq:E6cosetspec}
\eeqa
arise as localized matter there. 
On the right side here and below, the {\bf 5} and the ${\bf \bar 5}$ are denoted interchangeably 
as a hypermultiplet.
We call such a codimension-two $E_6$ singularity 
(enhanced from $SU(5)$) made of two {\bf 10}'s and one {\bf 5} a {\em complete 
singularity}, whereas one made of a single {\bf 10} and a single {\bf 5} an {\em incomplete 
singularity}\footnote{In \cite{MT}, the terms {\em complete/incomplete resolutions} were 
used; we will slightly change the nomenclature 
because what is incomplete is not the process of resolution and so it is somewhat misleading.}.
(In this paper, we ignore the match of the number of singlets, focusing only the 
change of the numbers of charged hypermultiplets.)

Similarly, we define a complete singularity in the enhancement 
to $G'=E_7$ and $E_8$ as one made up of generic codimension-two singularities 
that generate the same set of charged hypermultiplets as $G'/(SU(5)\times U(1)^{{\rm rank}G'-4})$.
If, on the other hand, an $G'=E_7$ or $E_8$ singularity made of a coalescence of 
generic singularities that support smaller number of hypermultiplets will be called 
an incomplete singularity.

In the cases where half-hypermultiplets are involved, 
the change in the structure of the 
singularity successfully explains the matter generation expected to occur there \cite{MT,halfhyper}.
So, then, in the case of multiple enhancement where ordinary hypermultiplets (that is, 
full hypermultiplets that are not half ones) gather, 
is there also a singularity structure that can account for the matter generation of that much?
This is the question that we would like to address in this paper.

\subsection{Incomplete singularities in $SU(5)\rightarrow E_6$, $E_7$ and $E_8$}
\label{section:incompleteE6E7E8}

As we saw in the previous subsection, how many hypermultiplets 
(or rather, how many generic codimension-two singularities that generate them) 
gather is determined by the order in $w$ of relevant sections that vanish there.
In the following, we will examine each case of $G'=E_6, E_7$, and $E_8$ in turn 
to see what they are in detail.

The anomaly-free localized charged matter spectrum at any multiply enhanced point is
given by
\beq
    {\mbox o}_w(a_{1,0})\cdot{\bf 10}  \oplus {\mbox o}_w(P_{8,5})\cdot{\bf 5}.
\label{eq:anomalyfreeHc}
\eeq
In the generic $SU(5)$ model, the anomaly-free spectrum is given by \eqref{eq:SU5Hc}
arised from $n+2$ $SO(10)$ points and $3n+16$ $SU(6)$ points. 
Suppose that a multiply enhanced point is the overlap of $p$ $SO(10)$ points and $q$ $SU(6)$ points.
Then, if the spectrum at the multiply enhanced point is merely the direct sum $p\cdot{\bf 10} \oplus q\cdot{\bf 5}$,
that is, if neither new representations arise nor any representations disappear through the enhancement,
it is obvious that the anomaly still cancels.
Since the degeneracies of the overlapping $SO(10)$ points and $SU(6)$ points at $w=0$ are
given by ${\mbox o}_w(a_{1,0})$ and ${\mbox o}_w(P_{8,5})$, respectively (see Table \ref{singorder}),
\eqref{eq:anomalyfreeHc} holds.
\\
\\
\noindent
\underline{$SU(5) \rightarrow E_6$}\\

As we have already seen in the  previous subsection, we can distinguish two cases (see Table \ref{E6pattern}).
\begin{table}[htbp]
\centering
\begin{tabular}{|ccccc|c|c|c|}
\hline
$\mbox{o}_w(a_{1,0})$  &  
$\hspace{-0.1cm}\mbox{o}_w(a_{2,1})$ &
$\hspace{-0.1cm}\mbox{o}_w(a_{3,2})$ &  
$\hspace{-0.1cm}\mbox{o}_w(a_{4,3})$ &
$\hspace{-0.1cm}\mbox{o}_w(a_{6,5})$ &  $\hspace{-0.0cm}\mbox{o}_w(P_{8,5})$& 
\mbox{spectrum} & \mbox{name}\\  \hline      
$1$&$1$&$0$&$0$&$0$&$1$& ${\bf 10}\oplus {\bf 5}$ & incomplete\\
$2$&$1$&$0$&$0$&$0$&$1$& $2\cdot{\bf 10}\oplus {\bf 5}$ & complete\\
\hline
\end{tabular}
\caption{$E_6$ patterns.}
\label{E6pattern}
\end{table}
%
%
Of course, we could consider further patterns in which more general codimension-two 
singularities overlap than in the complete case, but 
we limit ourselves to these cases in this paper.
Also, we note that the orders of $P_{8,5}$ are calculated for 
generic sections $a_{1,0},\ldots,a_{6,5}$ with the specified orders; they can 
be accidentally larger than them if $a_{1,0},\ldots,a_{6,5}$ satisfy some relation.
\\
\\
\noindent
\underline{$SU(5) \rightarrow E_7$}\\

If a set of exceptional curves occurs such that the intersection diagram coincides with
the complete $E_7$ Dinkin diagram, then the resulting hypermultiplets will be
\beqa
E_7/(SU(5)\times U(1)^3)&=&3\cdot {\bf 10}\oplus 4\cdot{\bf 5}\oplus 3\cdot{\bf 1},
\eeqa
where, again, ${\bf 5}$ and ${\bf \bar 5}$ are identified as the same hypermultiplet. 
Thus we define a complete $E_7$ singularity  as 
an $E_7$ singularity where $\mbox{o}_w(a_{1,0})=3$, $\mbox{o}_w(P_{8,5})=4$, 
and $\mbox{o}_w(a_{2,1})$, $\mbox{o}_w(a_{3,2})$, 
$\mbox{o}_w(a_{4,3})$ and $\mbox{o}_w(a_{6,5})$ take minimum values.
We also define various incomplete $E_7$ singularities as ones with 
$1\leq \mbox{o}_w(a_{1,0})\leq 3$ and $2\leq\mbox{o}_w(P_{8,5})\leq 4$ 
such that  $\mbox{o}_w(a_{2,1})$, $\mbox{o}_w(a_{3,2})$,  
$\mbox{o}_w(a_{4,3})$ and $\mbox{o}_w(a_{6,5})$ have minimum values for 
each fixed pair of $\mbox{o}_w(a_{1,0})$ and $\mbox{o}_w(P_{8,5})$.
Four such incomplete singularities can be found, and they are listed in Table 3 
together with the complete singularity.\\

\begin{table}[htbp]
\centering
\begin{tabular}{|ccccc|c|c|c|}
\hline
$\mbox{o}_w(a_{1,0})$  &  
$\hspace{-0.1cm} \mbox{o}_w(a_{2,1})$ &
$\hspace{-0.1cm} \mbox{o}_w(a_{3,2})$ & 
$\hspace{-0.1cm} \mbox{o}_w(a_{4,3})$ &
$\hspace{-0.1cm} \mbox{o}_w(a_{6,5})$ &  
$\mbox{o}_w(P_{8,5})$&  \mbox{spectrum} & \mbox{name} \\  \hline      
$1$&$1$&$1$&$0$&$0$&$2$& ${\bf 10}\oplus 2\cdot {\bf 5}$ & incomplete 1\\
$2$&$1$&$1$&$0$&$0$&$3$& $2\cdot{\bf 10}\oplus 3\cdot {\bf 5}$ & incomplete 2\\
$2$&$1$&$2$&$0$&$0$&$4$& $2\cdot {\bf 10}\oplus4\cdot  {\bf 5}$ & incomplete 3\\
$3$&$1$&$1$&$0$&$0$&$3$& $3\cdot {\bf 10}\oplus3\cdot {\bf 5}$ & incomplete 4\\
$3$&$2$&$1$&$0$&$0$&$4$& $3\cdot {\bf 10}\oplus4\cdot {\bf 5}$ & complete\\
\hline
\end{tabular}
\caption{$E_7$ patterns.}
\label{E7pattern}
\end{table}

\noindent
\underline{$SU(5) \rightarrow E_8$}\\

We can similarly list 45 patterns of complete and incomplete singularities.
In fact, we will see later that the details of Tate's orders for each pattern are not 
very relevant to the analysis of the resolution.
Since it is not very informative, we leave the results to Appendix A.

In the $E_8$ case
\beqa
E_8/(SU(5)\times U(1)^4)&=&5\cdot {\bf 10}\oplus 10\cdot{\bf 5}\oplus 10\cdot{\bf 1},
\eeqa
so a complete $E_8$ singularity should have $\mbox{o}_w(a_{1,0})=5$ and 
$\mbox{o}_w(P_{8,5})=10$. 
There are five such patterns, all 
named complete singularities in Table 4 in Appendix A.   

\section{Resolution of  multiply enhanced singularities in 6D F-theory I : Esole-Yau resolution}
\label{sec:EY}
\subsection{Generalities of the resolution of 6D $SU(5)$ models}
\label{subsection:Generalities}
In this section, we summarize the general aspects of singularity resolution 
in 6D $SU(5)$ F-theory models. This also enables us to state the results of \cite{EsoleYau} on 
the structure of the $SU(5)$ models in our notation, with some appropriate modifications  
to six dimensions.  

As we said, 
we work on an elliptic CY threefold over a Hirzebruch surface 
(\ref{eq:Tate0}).
By moving the terms on the left-hand side, let us write the equation as
\beqa
  \Phi(x,y,z,w) &\equiv &
    -  (y^2 + a_{1,0} x y +a_{3,2} z^2 y)+  x^3 + a_{2,1} z x^2 + a_{4,3}z^3 x +a_{6,5} z^5~=~0.
\label{Phi}
\eeqa
$(x,y,z,w)=(0,0,0,\mbox{any})$ is the codimension-one $SU(5)$ singularity 
at {\em generic} $w$. 
Let ${\mathfrak p}_0$ denote this singularity. 
As is well known, ${\mathfrak p}_0$ is desingularized by two-time insertions of 
lines of $\PP^2$'s  along $(x,y,z)=(0,0,0)$ with arbitrary $w$.

\bigskip

\noindent
\underline{Blow up of ${\mathfrak p}_0$}
\smallskip

The first insertion is done by setting\footnote{To be completely precise, (\ref{1stblowup})
is a particular expression of the $\PPsub^2$ blow-up 
$(x,y,z)=(\delta_1 X_1, \delta_1 Y_1, \delta_1 Z_1)$, $(X_1:Y_1:Z_1)\in\PPsub^2$ 
in the patch $Z_1\neq 0$, where $(x_1:y_1:1)=(X_1:Y_1:Z_1)$ are (the former) the affine 
coordinates in this patch, and $z$ itself becomes the variable for 
projectivization $\delta_1$.  Similarly, (\ref{2ndblowup}) is the 
expression of the blow-up $(x_1,y_1,z)=(\delta_2 X_2, \delta_2 Y_2, \delta_2 Z_2)$, $(X_2:Y_2:Z_2)\in\PPsub^2$
in the patch $X_2\neq 0$. 
}
\beqa
(x,y,z)&=&(x_1 z, y_1 z, z).
\label{1stblowup}
\eeqa
By plugging (\ref{1stblowup}) into (\ref{Phi}), we define 
\beqa
\Phi_z(x_1,y_1,z,w) &\equiv &z^{-2}  \Phi(x_1 z,y_1 z,z,w)\n
&=&    -  y_1(y_1 + a_{1,0} x_1 +a_{3,2} z)+  z(x_1^3 + a_{2,1} x_1^2 + a_{4,3}z x_1 +a_{6,5} z^2)=0,\n
\label{Phiz}
\eeqa
which is called the proper transform. Note that by factoring out $z^{2}$, the 
canonical class is preserved so that the new threefold $\Phi_z(x_1,y_1,z,w)=0$ remains 
a Calabi-Yau.
By setting $z=0$ on $\Phi_{z}=0$, we obtain the exceptional set 
\beq
\begin{cases}
C^{+}_1 \,:\, z=0 , \, \, y_1 = 0 , \\
C^{-}_1  \,:\, z=0 , \, \, y_1 + a_{1,0} x_1 = 0
\end{cases}
\label{eq:E6incompEYC1}
\eeq
along $w$.
The proper transform $\Phi_{z}=0$ is singular at $(x_1,y_1,z,w)=(0,0,0,\mbox{any})$.
Let ${\mathfrak p}_1$ denote this codimension-one singularity.

\bigskip

\noindent
\underline{Blow up of ${\mathfrak p}_1$}
\smallskip

For the second insertion of a line of $\PP^2$, we use 
\beqa
(x_1,y_1,z)&=&(x_1, x_1 y_2, x_1 z_2)
\label{2ndblowup}
\eeqa
in (\ref{Phiz}) to obtain
\beqa
\Phi_{zx}(x_1,y_2,z_2,w) &\equiv &x_1^{-2}  \Phi_z(x_1, x_1y_2 , x_1 z_2,w)\n
&=&     -  y_2(y_2 + a_{1,0} +a_{3,2} z_2)+  x_1z_2(x_1 + a_{2,1}  + a_{4,3}z_2  +a_{6,5} z_2^2)
=0.\n
\label{Phizx}
\eeqa
Again, factoring out $x_1^{2}$ yields a ``crepant'' resolution, meaning that it does not
change the canonical class.
The exceptional set is
\beq
\begin{cases}
C^{+}_2 \,:\, x_1=0 , \, \, y_2 = 0 , \\
C^{-}_2  \,:\, x_1=0 , \, \, y_2 + a_{1,0} + a_{3,2} z_2 = 0
\end{cases}
\label{eq:E6incompEYC2}
\eeq
for arbitrary $w$.
For generic $w$, the exceptional set consists of four components $C_i^{\pm}$ ($i=1,2$).
Their intersection forms $A_4$ Dynkin diagram, yielding the $SU(5)$ gauge symmetry
(see Figure \ref{Fig:E6incompEY} below).

After this second blow-up, 
the new threefold defined by the equation $\Phi_{zx}(x_1,y_2,z_2,w)=0$ is regular 
except for codimension-two discrete loci on the base, on the fibers 
over which conifold singularities appear \cite{EsoleYau, MT}.  This can be clearly 
seen by rewriting (\ref{Phizx}) as
\beqa
u_1u_2u_3-v_1v_2&=&0 \label{v1v2-u1u2u3=0again}
\eeqa
with
\beqa
u_1&=&x_1,\n
u_2&=&z_2,
\n
u_3&=&x_1 + a_{2,1}  + a_{4,3}z_2  +a_{6,5} z_2^2.\n
v_1&=&y_2,
\n
v_2&=&y_2 + a_{1,0} +a_{3,2} z_2.
\eeqa
We have already shown these equations as (\ref{v1v2-u1u2u3=0}) in Introduction. 

From (\ref{v1v2-u1u2u3=0again}), we can see that there are three 
types of conifold singularities in a generic six-dimensional 
$SU(5)$ model:
\begin{itemize}
\item{$v_1=v_2=0$ and $u_1=u_2=0$}\\
This occurs if $a_{1,0}=0$. In this case $y_2=x_1=z_2=0$.
In generic cases where $a_{2,1}\neq 0$, $u_3$ does not vanish. 
We call this conifold singularity ${\mathfrak v}_2$.
\item{$v_1=v_2=0$ and $u_2=u_3=0$}\\
This also occurs if $a_{1,0}=0$. In this case $y_2=z_2=0$ and $x_1=-a_{2,1}$.
Again, in generic cases where $a_{2,1}\neq 0$, 
$(x_1=)\,u_1$ does not vanish, so this is (generically) 
a different conifold singularity than ${\mathfrak v}_2$. 
We call this ${\mathfrak v}_1$.
\item{$v_1=v_2=0$ and $u_1=u_3=0$}\\
This type of conifold singularity occurs if $z_2$ can simultaneously satisfy
$a_{1,0} +a_{3,2} z_2=0$ and $a_{2,1}  + a_{4,3}z_2  +a_{6,5} z_2^2=0$.
This is when $P_{8,5}$ (\ref{eq:PQR}) is $0$. In this case $y_2=x_1=0$, and 
$z_2$ is a common solution to the two equations. We call this conifold singularity 
${\mathfrak u}_2$. 
\end{itemize}

The first two arise if $a_{1,0}=0$; they are the conifold singularities 
responsible for the generation of a {\bf 10} hypermultiplet at codimension-two $SO(10)$ points 
on the base. Similarly, the last one appears if $P_{8,5}=0$, so is the one 
that generates a {\bf 5} hypermultiplet. ${\mathfrak v}_1$ can already be seen in the locus of 
$\Phi_z$ (\ref{Phiz}), hence the name with the index ``1''.

\bigskip

\noindent
\underline{Esole-Yau resolution}
\smallskip

As stated above, these conifold singularities are all distinct unless
$a_{1,0}$ and $a_{2,1}$ simultaneously vanish, and then they are all desingularized 
by two additional small resolutions \cite{EsoleYau}. 
This is done by taking two pairs of sections $(v_1,u_i)$, $(v_2,u_j)$ 
$(i,j=1,2,3)$ and considering the projectivizations
\beqa
v_1=\xi V_1,~~~u_i=\xi U_i,
\label{EYsmallresolution1general}\\
v_2=\zeta V_2,~~~u_j=\zeta U_j,
\label{EYsmallresolution2general}
\eeqa
where $(V_1: U_i)$, $(V_2: U_j)$ are sections of $\PP^1$ bundles.
Specifically, if we take $i=1$ and $j=2$ for instance, 
(\ref{EYsmallresolution1general})
(\ref{EYsmallresolution2general})
reduce
to 
(\ref{EYsmallresolution1})
(\ref{EYsmallresolution2}), or
\beqa
y_2=\xi V_1,~~~x_1=\xi U_1,
\label{EYsmallresolution1specific}\\
y_2 + a_{1,0} +a_{3,2}z_2=\zeta V_2,~~~z_2=\zeta U_2.
\label{EYsmallresolution2specific}
\eeqa
Doing these replacements in (\ref{Phizx}) and factoring out $\xi\zeta$, 
we obtain
\beqa
\Phi_{zx\xi\zeta}((V_1:U_1),\xi,(V_2:U_2),\zeta
) &\equiv &\xi^{-1}\zeta^{-1}  
\Phi_{zx}\left(\xi U_1, \xi V_1, \zeta U_2, w((V_1:U_1),\xi,(V_2:U_2),\zeta)
\right) \n
&=&     -  V_1 V_2+  U_1 U_2\left(\xi U_1 + a_{2,1} (w) + a_{4,3}(w)\zeta U_2  +a_{6,5}(w) (\zeta U_2)^2\right)\n
&=&0,
\label{Phizx34}
\eeqa
where $w((V_1:U_1),\xi,(V_2:U_2),\zeta)$ is the implicit function determined by 
the first equation of (\ref{EYsmallresolution2specific}), or
\beqa
\xi V_1 + a_{1,0}(w) +a_{3,2}(w) \zeta U_2=\zeta V_2.
\label{constraint}
\eeqa
%
Later we will see that this constraint equation plays a significant role in the 
analysis of the structure of multiply enhanced singularities.

Since (\ref{EYsmallresolution1specific}) (or (\ref{EYsmallresolution1})) is a small resolution 
for the conifold singularities ${\mathfrak v}_2$ and ${\mathfrak u}_2$, and 
(\ref{EYsmallresolution2specific}) (or (\ref{EYsmallresolution2})) is 
for  ${\mathfrak v}_2$ and ${\mathfrak v}_1$, all the conifold singularities in a generic $SU(5)$ model are
resolved by these two small resolutions. 
Note that, although both 
(\ref{EYsmallresolution1specific})
and 
(\ref{EYsmallresolution2specific}) 
are small resolutions for ${\mathfrak v}_2$, only one (and not two) $\PP^1$('s) is inserted here 
because 
$(V_1,U_1)$ and $(V_2,U_2)$ are not independent but constrained by (\ref{constraint}).

Matter representations are extracted from the resolved geometry as follows \cite{MT,halfhyper}. 
First, the set of the exceptional curves at the codimension-two locus are identified.
Let $\delta_i$ denote these curves.
Some of them are the $w=0$ slice of the generic components $C_i$'s  
and the others are the $\PP^1$'s inserted via the small resolution that are existing only at $w=0$.
Next, the decomposition of each $C_i$ into $\delta_i$'s at $w=0$ is determined.
Then, the intersection matrix among $\delta_i$'s is calculated and from this matrix 
one can read the representations existing at $w=0$.
As the simplest example, we briefly show the result for the ``Esole-Yau-resolved'' $E_6$ incomplete singularity.
In this case, $o_w(a_{1,0}) = o_w(a_{2,1})=1$ (see Table \ref{E6pattern}).
From \eqref{eq:E6incompEYC1} and \eqref{eq:E6incompEYC2}, one can see that $C_1^{+}$ and $C_1^{-}$
coincide at $w=0$ to be a single exceptional curve $\delta_1$, whereas $C_2^{+}$ and $C_2^{-}$ 
yield two curves $\delta_2^+$ and $\delta_2^-$ at $w=0$.
There are two more $\PP^1$'s at $w=0$ arised from the Esole-Yau resolution.
We call them $\delta_{\xi}$ and $\delta_{\zeta}$ for the small resolutions \eqref{EYsmallresolution1specific} and
\eqref{EYsmallresolution2specific}, respectively.
These five $\delta$'s intersect as shown in Figure \ref{Fig:E6incompEY}.
\begin{figure}[htb]
  \begin{center}          
         \includegraphics[clip, width=12.2cm]{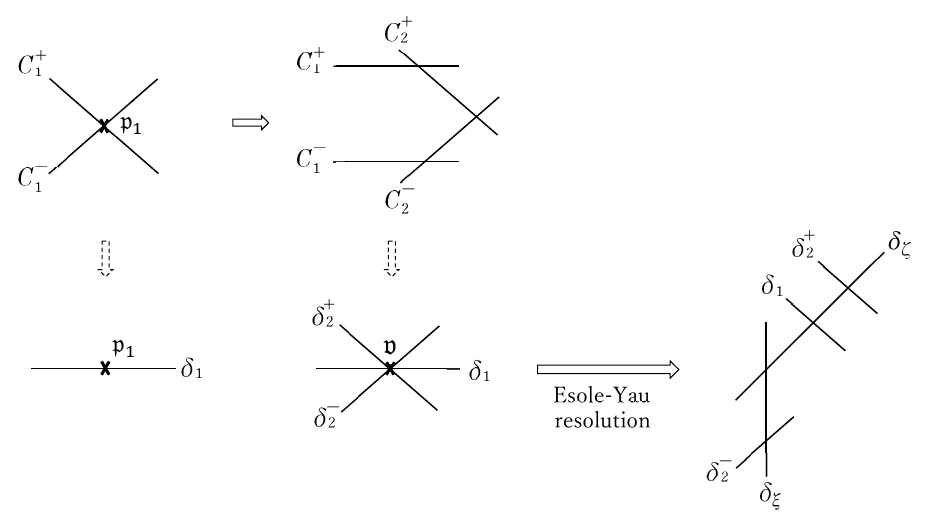}
       \caption{An example of the Esole-Yau resolution (incomplete $E_6$). ${\mathfrak p}_1$ is a codimension-one singularity; 
      ${\mathfrak v}$ is a codimension-two singularity. The down arrows denote the limit $w\rightarrow 0$. 
        }
    \label{Fig:E6incompEY}
  \end{center}
\end{figure}
Calculating the intersection matrix among 
these $\delta$'s and extracting the spectrum from that matrix\footnote{The method of these calculations 
will be described in detail in the next section.
Following the method, we obtain the intersection matrix for the five $\delta$'s.
It is the same matrix as presented in \eqref{eq:E6incompdeltaCartan} 
(see also Figure \ref{Fig:E6incompCartan}). 
$\delta_{\mathfrak v_2}$ and $\delta_{\mathfrak v_3}$ there are replaced to 
$\delta_{\xi}$ and $\delta_{\zeta}$ here.
Correspondingly, the resulting representations are same as \eqref{eq:E6incompspec}. },
one can show that $\bold{10} \oplus \bold{5}$ are existing at the incomplete $E_6$ point.
Since five $\delta_i$'s are short of forming the $E_6$ Dynkin diagram, 
the spectrum is smaller than the complete one ($2\cdot {\bf 10}\oplus {\bf 5}$),
but it is exactly what we expected from the anomaly cancellation for incomplete $E_6$.
As seen in Table \ref{E6pattern},\,\ref{E7pattern} and \ref{E8pattern}, a worse singularity 
gives a larger set of anomaly-free representations at $w=0$.
This is presumably because the worse singularity gives the larger 
exceptional set $\{ \delta_i \}$ at $w=0$.
Whether this is true will be examined hereafter.

\subsection{Esole-Yau resolution of multiply enhanced singularities: The first look}
\label{thefirstlook}
So far, we have given an overview of the generalities of the singularity resolution 
in a generic 6D $SU(5)$ F-theory model.  In fact, since we have discussed 
the Esole-Yau resolutions quite generally, we can just use the formula of the 
proper transform (\ref{Phizx34}) to consider multiply enhanced singularities 
if we assume that the various sections in Tate's form have the designated orders in $w$ 
shown in Tables \ref{E6pattern},\,\ref{E7pattern} and \ref{E8pattern}. For example, 
one can have the equation for the ``Esole-Yau-resolved'' $E_6$ incomplete singularity 
by setting the orders of $(a_{1,0}, a_{2,1}, a_{3,2}, a_{4,3}, a_{6,5})$ in $w$ to 
$(1,1,0,0,0)$ in (\ref{Phizx34}), 
and also have the one for the ``Esole-Yau-resolved'' $E_6$ complete singularity 
by setting them to $(2,1,0,0,0)$, respectively\footnote{More precisely, 
the resulting smooth model obtained by this choice of pairs is called ${\mathscr  B}_{2,1}$ 
in \cite{EsoleShaoYau1,EsoleShaoYau2} among the six Esole-Yau resolutions.
}.
Here, however, we are faced with a somewhat puzzling fact: Since $\Phi_{zx\xi\zeta}$ 
(\ref{Phizx34}) is 
of the form
\beqa
\Phi_{zx\xi\zeta}&=& -  V_1 V_2+  U_1 U_2 \left(\xi U_1 +\cdots \right),
\eeqa
\begin{itemize}
\item[(i)]{We can set $V_1 =U_2=1$ in the patch $V_1 \neq 0 ~\bigcap~ U_2 \neq 0$  to read
\beqa
\Phi_{zx\xi\zeta}&=& -  V_2+\cdots.
\label{Phizx34(i)}
\eeqa}
\item[(ii)]{We can set $V_2=U_1 =1$ in the patch $V_2\neq 0~\bigcap~ U_1 \neq 0$ to read
\beqa
\Phi_{zx\xi\zeta}&=& -  V_1+\cdots.
\label{Phizx34(ii)}
\eeqa}
\item[(iii)]{We can set $V_1 =V_2=1$ in the patch $V_1 \neq 0~\bigcap~ V_2\neq 0$ to read
\beqa
\Phi_{zx\xi\zeta}&=& -  1+\cdots.
\label{Phizx34(iii)}
\eeqa}
\item[(iv)]{We can set $U_1=U_2=1$ in the patch $U_1 \neq 0~\bigcap~ U_2\neq 0$ to read
\beqa
\Phi_{zx\xi\zeta}&=& - V_1 V_2 + \xi +\cdots.
\label{Phizx34(iv)}
\eeqa}
\end{itemize}
In all these cases, the equation $\Phi_{zx\xi\zeta}=0$ appears to be regular 
no matter how high the orders of the sections are, as long as $((V_1:U_1),\xi,(V_2:U_2),\zeta)$ 
are the coordinates!
%

In fact, this argument is too na\"{i}ve, and careful consideration will show 
in the later section that this last ``proviso'' is no longer valid.
But before we consider this, we will discuss in the next section 
an alternative way of small resolutions of multiply enhanced singularities. 

\section{Resolution of  multiply enhanced singularities in 6D F-theory II : Alternative
small resolution}
\label{Alternativesmallresolution}
\subsection{Alternative small resolution}
In this section we consider the small resolutions 
\eqref{Tanismallresolution1}\,\eqref{Tanismallresolution2}, or more specifically 
\beqa
y_2=\xi V_1,~~~x_1=\xi U_1,
\label{Tanismallresolution1specific}\\
V_1=\eta V'_1,~~~
\xi U_1 + a_{2,1}  + a_{4,3}z_2  +a_{6,5} z_2^2=\eta U_3.
\label{Tanismallresolution2specific}
\eeqa

As we already mentioned in Introduction, this way of small resolution 
(\ref{Tanismallresolution2specific}) is an 
equivalent change of the center of the blow up for ordinary conifold singularities, 
and so if there were no multiply enhanced singularities, it should have been 
classified as the same smooth model ${\mathscr  B}_{2,3}$\footnote{
in the notation in \cite{EsoleShaoYau1,EsoleShaoYau2}. 
Also, using the same notation, the model obtained by 
the blow up (\ref{Tanismallresolution2}) can be represented as
\beqa
{\mathscr  B}_{2,\bullet}\stackrel{(y,t|e_4)}{\longleftarrow}\tilde{\mathscr  B}_{2,3},   \nonumber
\eeqa
where we have indicated the smooth model with a tilde to 
distinguish it from ${\mathscr  B}_{2,3}$. 
In fact, although the proper transforms of the threefold equations are
indeed different between ${\mathscr  B}_{2,3}$ and 
$\tilde{\mathscr  B}_{2,3}$,  they
turn out to be ultimately equivalent when viewed from a certain perspective, 
which will be discussed in section \ref{duality}.
}.
Using these equations in 
(\ref{Phizx}) and dividing it by $\xi\eta$, we can derive  in the $U_1=V'_1=1$ patch 
(which is the only relevant one)
\beqa
\Phi_{zx\xi\eta}(U_3,\eta,z_2,w
) &\equiv &\xi^{-1}\eta^{-1}  
\Phi_{zx}\left(\xi(U_3,\eta,z_2,w) 
, \xi(U_3,\eta,z_2,w) \eta V'_1, z_2, w\right) \n
&=&    z_2(U_3-a_{3,2}(w)) -\xi(U_3,\eta,z_2,w) \eta - a_{1,0}(w) \n
&=&0,
\label{Phizxxieta}
\eeqa
where 
\beqa
\xi(U_3,\eta,z_2,w)&\equiv&\eta U_3-
(
a_{2,1}(w)  + a_{4,3}(w)z_2  +a_{6,5}(w) z_2^2
).
\label{xiconstraint}
\eeqa
Unlike $\Phi_{zx\xi\zeta}$, which always accompanies a first-order (or constant) term 
(\ref{Phizx34(i)})-(\ref{Phizx34(iv)}), 
these formulas show that new singularities can appear depending 
on the orders of the sections in Tate's form.
In the following, we will examine the structure of the singularity 
in detail in each case of enhancement to $E_6$, $E_7$, and $E_8$.

\subsection{$SU(5) \rightarrow E_6$}
\label{sec:E6}
\subsubsection{Incomplete $E_6$}
\label{sec:incompE6}

Let us first consider the incomplete $E_6$ singularity. 
To focus on such a specific point, we set $a_{1,0} =w$, 
where the coefficient is set to $1$ without loss of generality.
Substituting 
\bes
   a_{1,0} & = w,\\
   a_{2,1} & = w\, c 
\label{eq:a10a21}  
\ees
into (\ref{Phi}), we obtain Tate's form of the geometry
\beq
  \Phi \equiv -(y^2 + w x y +  a_{3,2} z^2 y ) +x^3 +w \, c \, z x^2+a_{4,3} z^3 x +a_{6,5} z^5 = 0.
\label{eq:E6incompPhi}
\eeq
It contains singularities aligned at $(x,y,z,w)=(0,0,0,w)$, which we denote as ${\mathfrak p}_0$. 

\bigskip

\noindent
\underline{Blow up of ${\mathfrak p}_0$}
\smallskip

\begin{figure}[hb]
  \begin{center}          
         \includegraphics[clip, width=15.6cm]{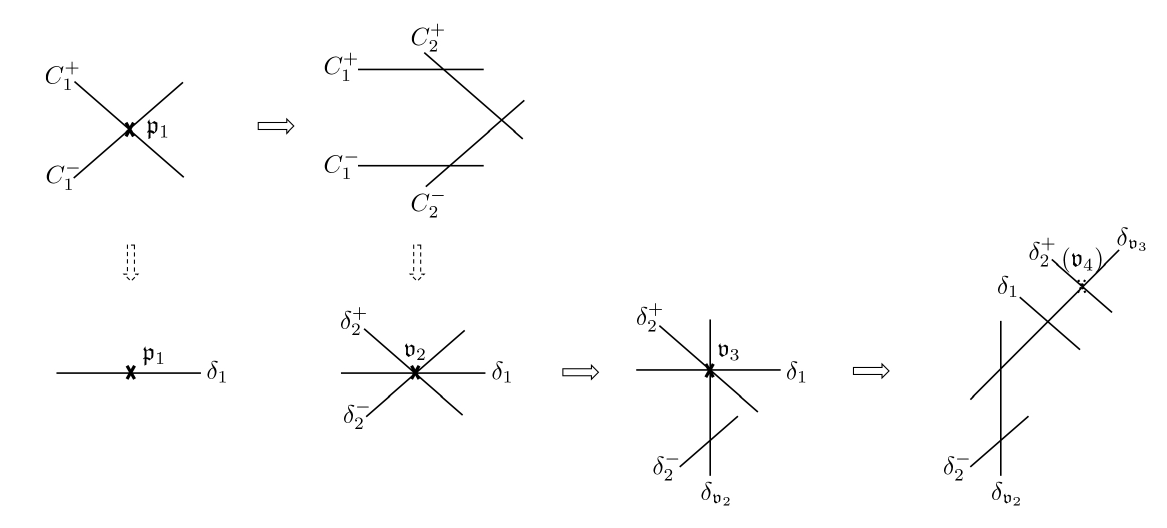}
       \caption{The resolution process of the incomplete $E_6$. ${\mathfrak p}_1$ is a codimension-one singularity; ${\mathfrak v}_2$ and ${\mathfrak v}_3$ are 
                   codimension-two singularities. The down arrows denote the limit $w\rightarrow 0$. 
                   (For the complete case, one more codimension-two singularity ${\mathfrak v}_4$ 
                   remains. See section \ref{sec:E6comp}.)}
    \label{Fig:E6incomp}
  \end{center}
\end{figure}

To blow up ${\mathfrak p}_0$, 
we set
$
(x,y,z) = (x_1 z,y_1 z,z)
$
as in (\ref{1stblowup}).
The geometry after this blow up 
is given by (\ref{Phiz}) with \eqref{eq:a10a21}:
\beq
\Phi_z  \equiv z^{-2} \Phi 
        =-y_1\{y_1+(w x_1 + a_{3,2} z)\} +z(x_1^3+w \, c \, x_1^2+a_{4,3} z x_1 +a_{6,5}z^2) =0.
\label{eq:E6incomp1zcurve}
\eeq
It still contains a codimension-one singularity at $(x_1,y_1,z,w)=(0,0,0,w)$, which we call ${\mathfrak p}_1$.
The exceptional sets $C^{\pm}_1$ aligned over the curve ${\mathfrak p}_0$ are obtained 
by setting $z=0$ in \eqref{eq:E6incomp1zcurve} 
and the exceptional curve $\delta_1$ at the $E_6$ 
point are given by taking their $w\rightarrow 0$ limit (Figure \ref{Fig:E6incomp}):
\beq
\begin{cases}
C^{+}_1 \,:\, z=0 , \, \, y_1 = 0  \\
C^{-}_1  \,:\, z=0 , \, \, y_1 = - w x_1 
\end{cases}
   \quad   \rightarrow   \quad    \delta_1 \, :\, w=0,\,\,z=0,\,\, y_1 = 0.  
\label{eq:E6incompzset}
\eeq

%
\bigskip

\noindent
\underline{Blow up of ${\mathfrak p}_1$}
\smallskip

To blow up ${\mathfrak p}_1$, we set 
$
 (x_1,y_1,z) = (x_1, x_1 y_2 , x_1 z_2) 
$
as in (\ref{2ndblowup}).
The 
geometric data after this blow up are given as follows:
\bes
& \Phi_{zx} \equiv x_1^{-2} \Phi_z 
     =  -y_2\{y_2+(w + a_{3,2} z_2)\} + x_1 z_2 (x_1+w \, c +a_{4,3}z_2 + a_{6,5}z_2^2) =0 .\\
& \mbox{Singularity} \,: \, {\mathfrak v}_2 = (0,0,0,0) ,   \\
& 
\begin{cases}
C_2^+ \,:\, x_1=0 ,\,\, y_2 = 0 \\
C_2^- \,:\, x_1=0 ,\,\, y_2 = - (w+a_{3,2} z_2)
\end{cases}
 \hspace{-0.1cm}  \rightarrow    \hspace{0.2cm}  
\begin{cases}
 \delta_2^+ \,:\, x_1=0 ,\,\, y_2=0,\,\, w= 0 \\ 
 \delta_2^-  \,:\, x_1=0 ,\,\, y_2 = -a_{3,2}z_2,\,\, w=0 
\end{cases}
\\
& 
\begin{cases}
C_1^{+}\,:\, z_2=0,\,\, y_2 = 0   \\
C_1^{-}\,:\,  z_2=0,\,\, y_2= -w 
\end{cases}
   \hspace{2.9cm}   
    \hspace{-0.6cm}      \hspace{0.4cm}  
    \delta_1\, :\, z_2=0,\, y_2=0,\, w=0.
\label{eq:E6incompzxCdelta}
\ees
Here $C_2^{\pm}$ are the exceptional sets at generic $w$ arised via the 
blow up of $\mathfrak{p}_1$ and
$\delta_2^{\pm}$ are their $w\rightarrow 0$ limits.
$C_1^{\pm}$ and $\delta_{1}$ are the lift-ups of the objects defined in \eqref{eq:E6incompzset}.

The intersections of these objects are depicted in the second column of Figure \ref{Fig:E6incomp}.
$C^\pm_i$'s form the $SU(5)$ Dynkin diagram.
Three $\delta$'s meet at one point, where the singularity ${\mathfrak v}_2$ is located.
%
This ${\mathfrak v}_2$ is not a conifold singularity since the second term of $\Phi_{zx}$ is 
of order three or higher.
This type of singularities are known as {\em generalized conifold singularities}
\cite{generalizedconifoldsingularities1,generalizedconifoldsingularities2}.
Although it is not an ordinary conifold singularity, 
 it can be regarded as 
consisting of two overlapping conifold singularities, which 
can be resolved by inserting two $\PP^1$'s successively as we will see below. 

For this, we rewrite $\Phi_{zx}$ as we did in (\ref{v1v2-u1u2u3=0again}) as 
\beq
   \Phi_{zx}(u_1,u_2,u_3,v_1)
   = u_1u_2u_3
   -v_1\{v_1+(w+a_{3,2}u_2)\}.
\label{eq:E6incompzx}
\eeq
with
\bes
   u_1 &\equiv x_1 ,\\
   u_2 &\equiv z_2, \\
   u_3 &\equiv x_1+w\, c+a_{4,3}z_2 + a_{6,5}z_2^2 , \\
   v_1 &\equiv y_2.
\label{eq:E6incompuiv1}
\ees
$w$ in the second term of  \eqref{eq:E6incompzx} is written using these coordinates as 
\beq
  w = \frac{1}{c} (u_3-u_1- a_{4,3} u_2-a_{6,5} u_2^2 ).
\label{eq:E6incompw}
\eeq
In the four-dimensional case, the corresponding geometry has the form 
$u_1 u_2 u_3-v_1 v_2=0$ and the 
five coordinates are all independent, whereas in the six-dimensional case 
the coordinate $v_2$ is not independent.

The exceptional sets \eqref{eq:E6incompzxCdelta} are written in these coordinates as 
\bes
&
\begin{cases}
\displaystyle C_1^{+} \, : \, u_2=0,\ v_1 =0,\, w=\frac{1}{c}(u_3-u_1) ,   \\
\displaystyle C_1^{-} \, : \, u_2=0,\, v_1 =-w,\, w=\frac{1}{c}(u_3-u_1) ,   
\end{cases}
  \hspace{0.6cm}     \delta_1 \, : \, u_2=0,\, v_1=0 ,\, u_3=u_1, \\
& 
\begin{cases}
\displaystyle C_2^{+} \, : \, u_1=0,\, v_1=0, \, w= \frac{1}{c} ( u_3-a_{4,3}u_2-a_{6,5}u_2^2 ) ,\,  \\
\displaystyle C_2^{-} \, : \, u_1=0,\, v_1 = -(w+a_{3,2}u_2),  \, w= \frac{1}{c} ( u_3-a_{4,3}u_2-a_{6,5}u_2^2 ) ,\, 
\end{cases}
\\
&  \hspace{4.0cm}
 \rightarrow    \hspace{0.2cm}    
\begin{cases}
  \delta_2^{+} \,:\,   u_1=0,\, v_1=0,  \, u_3=a_{4,3}u_2+a_{6,5}u_2^2,\,  \\
  \delta_2^{-} \, :\,  u_1=0,\, v_1=-a_{3,2}u_2, \, u_3=a_{4,3}u_2+a_{6,5}u_2^2.
\end{cases}
\label{eq:E6incompzxCdelta2}
\ees

\bigskip

\noindent
\underline{Resolution of ${\mathfrak v}_2$}
\smallskip

For the first step of our resolution, 
we first insert $\PP^1$ in the $(u_1,v_1)$ plane as
\beq
  (u_1,v_1) \equiv (\xi U_1, \xi V_1)
\eeq
similarly to the Esole-Yau. 
There are two local coordinate patches: $U_1=1$ and $V_1=1$, which we call patch $1$ and patch $2$, respectively.
In patch $1$,
\beq 
(u_1,v_1)=(\xi,\xi V_1).
\label{eq:E6incompEY1}
\eeq
The resolved geometry is given by
\beq
  \Phi_{zx1}(\xi,u_2,u_3,V_1)  \equiv \xi^{-1} \Phi_{zx}(\xi,u_2,u_3,\xi V_1)  
                                        =u_2 u_3-V_1 (\xi V_1+w+a_{3,2}u_2) =0  
\label{eq:E6incompEY1curve}
\eeq
with 
\beq
w  = \frac{1}{c} (u_3-\xi-a_{4,3}u_2-a_{6,5}u_2^2) .
\eeq
It contains a residual codimension-two singularity at the origin 
which we call ${\mathfrak v}_3$:
\beq
     {\mathfrak v}_3 = (0,0,0,0).
\eeq
The inserted $\PP^1$ (denoted as $\delta_{{\mathfrak v}_2}$) corresponds to the original singularity ${\mathfrak v}_2=(u_1,u_2,u_3,v_1)=(0,0,0,0)$.
In this patch, it is
\beq
   \delta_{{\mathfrak v}_2} \,:\,  \xi=0,\, u_2 =0,\, u_3=0.
\label{eq:E6incompEY1deltav2}
\eeq
The 
exceptional sets \eqref{eq:E6incompzxCdelta2}  are now given by
\bes
&
\begin{cases}
\displaystyle C_1^+\, :\, u_2=0,\, V_1=0, \, w=\frac{1}{c}(u_3-\xi),   \\
\displaystyle C_1^-\, :\, u_2=0,\, \xi V_1=-w, \, w=\frac{1}{c}(u_3-\xi),
\end{cases}
\hspace{0.5cm} 
\hspace{0.2cm}      \delta_1\, : \, u_2=0,\, V_1=0,\, u_3=\xi,   \\
& 
\begin{cases}
\displaystyle C_2^+\, :\, \xi=0,\, u_2 u_3=V_1(w+a_{3,2}u_2), \, w=\frac{1}{c} ( u_3-a_{4,3}u_2-a_{6,5}u_2^2),  \\
\displaystyle C_2^-\, :\, \mbox{invisible},
\end{cases}
\\
&   \hspace{5.0cm}  
\begin{cases}
\delta_2^+ \, : \,  \xi=0,\, u_3 = a_{3,2}V_1,\,  u_3=a_{4,3}u_2+a_{6,5}u_2^2,  \\
\delta_2^- \,: \, \mbox{invisible}.
\end{cases}   
\label{eq:E6incompEY1exceptionalset}    
\ees
Three exceptional curves $\delta_{{\mathfrak v}_2}$, $\delta_1$ and $\delta_2^+$ are intersecting 
each other at one point and ${\mathfrak v}_3$ is located there 
(see the third diagram in the lower row in Figure \ref{Fig:E6incomp})\footnote{The equations \eqref{eq:E6incompEY1exceptionalset} are derived
by substituting \eqref{eq:E6incompEY1} into \eqref{eq:E6incompzxCdelta2}
and picking up the multiplicity-one independent component in $\Phi_{zx1}=0$.
For $C_1^+$, the substitution gives the form
\beq
 u_2=0,\,\,\, \xi V_1 =0,\,\,\, w=\frac{1}{c}(u_3-\xi).  \nonumber
\label{eq:E6incompzx1C1p}
\eeq
Since this 
$C_1^+$ is a subvariety of $\Phi_{zx1}=0$, an additional constraint 
$
 V_1 w =0 
$
should be satisfied. 
It is solved by $V_1=0$ since $w\neq 0$ at generic point of the codimension-one discriminant locus.
Then the second equation $\xi V_1=0$ is satisfied,
giving the form of $C_1^+$ in \eqref{eq:E6incompEY1exceptionalset}.
Also, $\delta_1$ is rewritten as  
\beq
  u_2=0,\,\,\, \xi V_1 = 0, \,\,\,  u_3=\xi,    \nonumber
\eeq
which satisfies $\Phi_{zx1}=0$.
It has two components $u_2=0, \xi=0,u_3=0$ and  $u_2=0, V_1=0,u_3=\xi$.
The former one is equivalent to $\delta_{{\mathfrak v}_2}$ \eqref{eq:E6incompEY1deltav2},
and hence the latter component gives the 
equation of $\delta_1$.
The forms of $C_1^-$, $C_2^+$ and $\delta_2^+$ are similarly obtained.
$C_2^-$ is given by the equations 
\beq
 \xi=0,\,\,\, 0=-(w+a_{3,2}u_2),\,\,\, w= \frac{1}{c}(u_3-a_{4,3}u_2-a_{6,5}u_2^2) , \,\,\, u_2 u_3 =0,  \nonumber
\eeq
where the last condition comes from $\Phi_{zx1}=0$.
From the second equation, $u_2\neq 0$ for generic $w$, and hence $u_3=0$ from the last equation.
The second and the third equations yield $w=-a_{3,2}u_2$ and  $-c\,a_{3,2}+a_{4,3}+a_{6,5}u_2=0$,
which have no solution for generic $w$.
Thus $C_2^-$ is invisible in this patch.
The 
equation of $\delta_2^-$ is equivalent to $\delta_{{\mathfrak v}_2}$ and has no independent component.}.

Na\"{i}vely, one might think of $\delta_1$ as arising from $C_1^+$ and $C_1^-$,
or $\delta_{\mathscr v_2}$ as from $C_2^+$ and $C_2^-$ only.
In fact, however, if one carefully examines how the defining equations of 
these exceptional sets factorize in the $w\rightarrow 0$ limit, 
one can recognize exactly which exceptional sets that occur in the limit 
are composed of which exceptional sets defined prior to taking the limit
\cite{MT,Yukawas,halfhyper}.
For example,
%
from \eqref{eq:E6incompEY1exceptionalset}, one can 
see that 
the limit $w\rightarrow 0$ of $C$'s are written in terms of $\delta$'s as
follows:
\beq
 C_1^+  
 \rightarrow 
 \delta_1, \quad 
 C_1^-  
 \rightarrow 
 \delta_1 + \delta_{{\mathfrak v}_2}, \quad
 C_2^+  
 \rightarrow 
 \delta_2^+ + \delta_{{\mathfrak v}_2}. 
\label{eq:E6incomplimit1}
\eeq
The limit of $C_2^+$ is obtained by
\bes
 \mbox{lim}_{w\rightarrow 0} \, C_2^+ &= \{\xi=0,\, u_2 u_3 =V_1 a_{3,2}u_2,\, u_3=a_{4,3}u_2+a_{6,5}u_2^2 \} \\
                                               &= \{ \xi=0,\, u_3=V_1 a_{3,2},\, u_3=a_{4,3}u_2+a_{6,5}u_2^2\} 
                                            \cup \{ \xi=0,\, u_2=0,\, u_3=0\} \\
                                               &= \delta_2^+ \cup \delta_{{\mathfrak v}_2}.
\label{eq:E6incompC2plimit}
\ees
%
Similar calculations will be done repeatedly throughout this paper.
Note that the summations in (\ref{eq:E6incomplimit1}) 
should be understood as those in the weight space of $SU(5)$ or divisors \cite{Yukawas}.

In the other patch $(u_1,v_1)=(\xi U_1,\xi)$  (patch 2, the $V_1=1$ patch),
the resolved geometry  is given by
\bes
 \Phi_{zx2}(U_1,u_2,u_3,\xi) & \equiv \xi^{-1} \Phi_{zx}(\xi U_1,u_2,u_3,\xi)  
   = U_1 u_2 u_3 - (\xi+w+a_{3,2} u_2) =0,   \\  
& \hspace{-0.3cm}   w  = \frac{1}{c}\big\{ u_3-\xi U_1-a_{4,3}u_2-a_{6,5}u_2^2 \big\}.
\ees
It contains no singularity and is regular.
The intersections and limits are similarly calculated as in patch $1$ 
(the $U_1=1$ patch). We found that 
$\delta_2^- \cdot \delta_{{\mathfrak v}_2} \neq 0 $
and ($C_1^+$ and $\delta_1$ are invisible)
\beq
  C_1^- 
 \rightarrow
\delta_{{\mathfrak v}_2} ,\quad 
  C_2^+
\rightarrow
 \delta_2^+ + \delta_{{\mathfrak v}_2},  \quad
  C_2^-
\rightarrow
 \delta_2^-.
\label{eq:E6incomplimit2}
\eeq
This completes the first step of our small
resolution. The result is summarized in the third diagram in the lower row of 
Figure \ref{Fig:E6incomp}.

The remaining codimension-two singularity ${\mathfrak v}_3$ 
in patch $1$ (the $U_1=1$ patch) is a conifold singularity 
since $\Phi_{zx1}$ \eqref{eq:E6incompEY1curve} has the form
\beq
 \Phi_{zx1}(\xi,u_2,u_3,V_1) = u_2 u_3  - V_1\{w +\cdots\}  = u_2 u_3  - V_1\big\{ -\frac{1}{c} \xi +\cdots \big\}.
\label{eq:E6incomp1conifold}
\eeq
It is resolved by a standard small resolution, which is the second step of 
our
resolution.
This means that ${\mathfrak v}_2$ consists of two overlapping conifold singularities.
\bigskip


\noindent
\underline{Resolution of ${\mathfrak v}_3$
}
\smallskip

\begin{figure}[htbp]
  \begin{center}          
         \includegraphics[clip, width=6.6cm]{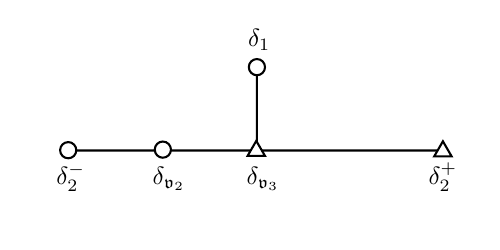}
       \caption{The intersection matrix of $\delta$'s for the incomplete $E_6$. The triangular nodes 
                    have self-intersection numbers $-\frac{3}{2}$ and their mutual intersection numbers are $\frac{1}{2}$. }
    \label{Fig:E6incompCartan}
  \end{center}
\end{figure}

For this small resolution, we choose the plane $(u_3,V_1)$ and insert a $\PP^1$ (denoted as $\delta_{{\mathfrak v}_3}$) as follows
:
\beq
  (u_3,V_1) = (\eta U_3, \eta V'_1).
\label{eq:u3V1}
\eeq
The two patches with $U_3=1$ and $V'_1=1$ are denoted as patch $1'$ and patch $2'$.
In patch $1'$ with $(u_3,V_1)=(\eta, \eta V'_1)$, the data of the resolved geometry is given by
\bes
& \Phi_{zx11'}(\xi,u_2,\eta,V'_1) \equiv \eta^{-1} \Phi_{zx1}(\xi, u_2,\eta,\eta V'_1)  
                                           =  u_2- V'_1(\xi \eta V'_1+w+a_{3,2} u_2)   = 0,  \\
& \hspace{2.7cm}
w =  \frac{1}{c} (\eta-\xi-a_{4,3}u_2 -a_{6,5}u_2^2).   \\
& \mbox{Singularity : none (regular)},  \\
& 
\begin{cases}
\displaystyle C_1^+\, :\, u_2=0,\, V'_1=0,\, w=\frac{1}{c}(\eta-\xi),    \\
\displaystyle C_1^-\, :\, u_2=0, \, \xi \eta V'_1=-w,\,  w=\frac{1}{c}(\eta-\xi), 
\end{cases} 
\hspace{1.5cm} 
 \delta_1\, : \, \, u_2=0, V'_1=0,\, \eta=\xi,  \\
&\hspace{0.4cm} C_2^+\, :\, \xi=0,\, u_2 = V'_1(w+a_{3,2} u_2),\, w=\frac{1}{c} ( \eta-a_{4,3}u_2-a_{6,5}u_2^2 ) ,   \\
 &   \hspace{6.5cm}  
 \delta_2^+ \, : \, \xi=0,\,  a_{3,2}V'_1=1,\, \eta=a_{4,3}u_2+a_{6,5}u_2^2, \\
&  \hspace{6.5cm}    
\delta_{{\mathfrak v}_2} \,:\, \mbox{invisible},\\
&  \hspace{6.5cm}     
\delta_{{\mathfrak v}_3} \,: \, \xi=0,\, \eta=0, u_2=0.   
\label{eq:E6incompzx11d}
\ees
$\delta_1$ and $ \delta_{{\mathfrak v}_3}$ intersect at the origin, whereas  $\delta_2^+$ and $ \delta_{{\mathfrak v}_3}$ intersect 
at another point $(\xi,u_2,\eta,V'_1)= (0,0,0,\frac{1}{a_{3,2}})$:
\beq
   \delta_1\cdot  \delta_{{\mathfrak v}_3} \neq 0, \quad  \delta_2^+\cdot  \delta_{{\mathfrak v}_3} \neq 0.
\label{eq:E6incomp11dintersection}
\eeq
The limits can be read from \eqref{eq:E6incompzx11d} as 
\beq
 C_1^+ 
 \rightarrow 
  \delta_1,\quad C_1^- 
 \rightarrow 
 \delta_1 + 2 \delta_{{\mathfrak v}_3},
\quad C_2^+ 
 \rightarrow 
 \delta_2^+ +  \delta_{{\mathfrak v}_3}.
\label{eq:E6incomplimit11d}
\eeq
For example,
\bes
 \mbox{lim}_{w\rightarrow 0}\, C_1^- &=  \{ u_2=0, \, \xi \eta V'_1 = 0, \, 0=\eta-\xi\}   \\
 & = \{u_2=0,\, \eta^2 V'_1=0, \xi = \eta\} \\
 & = \{ u_2=0,\, V'_1 =0,\, \xi=\eta\} \cup \{u_2=0,\, \eta=0,\, \xi=0\}^{\otimes 2}  \\
 & = \delta_1 \cup 2 \delta_{{\mathfrak v}_3}.
\ees
In patch $2'$ with $(u_3,V_1)=(\eta U_3,\eta)$ (the $V'_1=1$ patch), the resolved geometry is given by
\bes
&   \Phi_{zx12'}(\xi,u_2,U_3,\eta) \equiv \eta^{-1} \Phi_{zx1}(\xi, u_2,\eta U_3,\eta)  
    =u_2 U_3-(\xi \eta+w+a_{3,2} u_2)=0,\\
& \hspace{2.7cm} w=\frac{1}{c}( \eta U_3-\xi-a_{4,3}u_2
                                            -a_{6,5}u_2^2).   
\label{eq:E6incompzx12dPhi}
\ees
These equations are the same as (\ref{Phizxxieta}) and (\ref{xiconstraint}) with $a_{1,0}=w$ and $a_{2,1}=w c$ substituted. $\Phi_{zx12'}$ is denoted by $\Phi_{zx \xi \eta}$ in (\ref{Phizxxieta})\footnote{  
The constraint is solved for $w$ here, whereas it is (formally) solved for $\xi$ in (\ref{xiconstraint}).
Of cource the geometry is the same.}.
The other geometric data are given by
\bes
& \hspace{0.4cm} \mbox{Singularity\,:\, none (regular)}, \\
& 
\begin{cases}
\displaystyle C_1^+\, :\, \mbox{invisible} , \\
\displaystyle C_1^-\, :\, u_2=0,\,  \xi \eta =-w, \, w=\frac{1}{c}(\eta U_3-\xi),
\end{cases}
\hspace{0.5cm} 
 \hspace{2.2cm}
\delta_1\, : \,  \mbox{invisible}   ,  \\
& \hspace{0.4cm }C_2^+\, :\, \xi=0,\, u_2 U_3= (w+a_{3,2}u_2),\, w=\frac{1}{c}( \eta U_3-a_{4,3}u_2-a_{6,5}u_2^2 ), \\
&   \hspace{5.5cm}  \delta_2^+ \, : \, \xi=0, \, U_3=a_{3,2}, \, \eta U_3= a_{4,3} u_2+a_{6,5}u_2^2,  \\
& \hspace {5.5cm}    \delta_{{\mathfrak v}_2}\,:\, \xi=0,\, u_2=0, \, U_3=0, \\
& \hspace{5.5cm}     \delta_{{\mathfrak v}_3}\,:\, \xi=0,\, u_2=0,\, \eta=0. 
\label{eq:E6incompEY12dcurve}
\ees
$\delta_{2}^+$ and $ \delta_{{\mathfrak v}_3}$ intersect at $(\xi,u_2,U_3,\eta)=(0,0,a_{3,2},0)$,
whereas $ \delta_{{\mathfrak v}_2}$ and $ \delta_{{\mathfrak v}_3}$ intersect at the origin:
\beq
 \delta_2^+\cdot  \delta_{{\mathfrak v}_3} \neq 0,\quad  \delta_{{\mathfrak v}_2}\cdot  \delta_{{\mathfrak v}_3}\neq 0.
 \label{eq:E6incompint12d}
\eeq
The limits are given by
\beq
 C_1^- 
 \rightarrow 
  \delta_{{\mathfrak v}_2} + 2  \delta_{{\mathfrak v}_3},\quad
 C_2^+ 
 \rightarrow 
 \delta_2^+ +  \delta_{{\mathfrak v}_2} +  \delta_{{\mathfrak v}_3}.
\label{eq:E6incomplimit12d}
\eeq
This completes the resolution process.
The whole intersecting pattern is shown in the rightmost diagram in Figure \ref{Fig:E6incomp}.
The limits of $C_i$'s are obtained by taking the union of the results of all patches \eqref{eq:E6incomplimit1}, \eqref{eq:E6incomplimit2}, \eqref{eq:E6incomplimit11d} and \eqref{eq:E6incomplimit12d} as
\bes
 C_1^+ & 
 =
   \delta_1 , \\
 C_1^- & 
 =
   \delta_1 +  \delta_{{\mathfrak v}_2} + 2 \delta_{{\mathfrak v}_3}, \\
 C_2^+ & 
 =
   \delta_2^+ +  \delta_{{\mathfrak v}_2} +  \delta_{{\mathfrak v}_3}, \\
 C_2^- & 
 =
   \delta_2^-,  
\label{eq:E6incomplimit}
\ees
where, again, the equalities should be understood as those of $SU(5)$ weights or divisors.  
Under this identification, one can easily check that the intersection matrix of $C_i$'s is equivalent 
to (minus) the $SU(5)$ Cartan matrix if the intersection matrix of $\delta_i$'s has the form
\beq
   \delta_i\cdot \delta_j 
 = - \left( 
\begin{array}{rrrrr}
  2    &      -1        &         0        &    0              &    0     \\
 -1   &       2         &       -1        &    0              &    0     \\
  0    &     -1         & \frac{3}{2}    &  -\frac{1}{2}  &  -1     \\
  0    &       0         & -\frac{1}{2}  &    \frac{3}{2}  &     0                \\
  0    &       0        &         -1        &    0             &     2     
\end{array} 
    \right),
\label{eq:E6incompdeltaCartan}
\eeq
where the rows and the columns are ordered as $\delta_2^-,  \delta_{{\mathfrak v}_2},  \delta_{{\mathfrak v}_3}, \delta_2^+$ and $\delta_{1}$.
This intersection matrix is depicted in Figure \ref{Fig:E6incompCartan}.
The second node from the right of the $E_6$ Dynkin diagram is removed\footnote{
If we insert $\PPsub^1$ in the plane $(u_2,V_1)$ 
instead of the plane $(u_3,V_1)$ \eqref{eq:u3V1} in the second 
step of the 
resolution, the intersection matrix is the one 
that the center node of the $E_6$ Dynkin diagram is removed.
\label{footnote:u2V1}}. 

Let us now define curves $J$ at the incomplete $E_6$ point as linear combinations of $\delta$'s with integer coefficients via
\beq
 J \equiv  \sum_{n_i} n_i \delta_i \quad  \mbox{with}  \quad n_i \in \ZZ.
 \label{eq:J}
\eeq  
The intersection matrix of $\delta$'s defines a lattice which is the projection of the root lattice of $E_6$
in the direction orthogonal to the root of the removed node.
By this projection, the curves corresponding to the roots of $E_6$, which have 
self intersection number $J.J=-2$, are projected to the states with $-2\leq J.J < 0$
(apart from $J.J=0$).
It is shown that possible such self-intersections of $J$ are $-2$ or $-\frac{3}{2}$ and the
numbers of such curves are $30$ or $20$, respectively.
These curves can be thought of as forming a representation of a charged matter hypermultiplet.
Former curves are ``adjoint of incomplete $E_6$'' and taking their coset by $SU(5)$ 
gives  $\bold{5}$, 
while the latter curves form $\bold{10}$: 
\bes
  & \sharp(J.J= -2) = 30  \rightarrow  \bold{5},  \\
  & \sharp(J.J=-\frac{3}{2})  = 20 \rightarrow  \bold{10}.
\label{eq:E6incompspec}
\ees
In conclusion, an incomplete $E_6$ singularity gives matter multiplet $\bold{10} \oplus \bold{5}$,
which is nothing but the expected result from the anomaly-free condition (see Table \ref{E6pattern}). 

\subsubsection{Complete $E_6$}
\label{sec:E6comp}

Next let us move on to  the complete $E_6$ geometry. 
Since 
$(\mbox{o}_w(a_{1,0}),\mbox{o}_w(a_{2,1}),\mbox{o}_w(a_{3,2}),\mbox{o}_w(a_{4,3}),\\
\mbox{o}_w(a_{6,5}),\mbox{o}_w(P_{8,5}))=(2,1,0,0,0,1)$ are the conditions 
for a complete $E_6$ singularity, we set
\bes
  a_{1,0} & = w^2 , \\
  a_{2,1} & = w \, c.
\ees
The 
equation (\ref{Phi}) of this geometry is given by 
\beq
  \Phi \equiv -(y^2 + \underline{w^2} x y +  a_{3,2} z^2 y ) +x^3 +w \,c \,z x^2+a_{4,3} z^3 x +a_{6,5} z^5 = 0.
\label{eq:E6compPhi}
\eeq
It differs from the incomplete case in the underlined term.

The first and second resolutions of codimension-one 
singularities proceed in the same manner as was done in the incomplete case.
The geometry after the resolution of ${\mathfrak p}_1$ is similar to the incomplete case \eqref{eq:E6incompzx} 
and has the form 
\beq
 \Phi_{zx} = u_1 u_2 u_3-v_1\{ v_1+(\underline{w^2}+a_{3,2}u_2)\} = 0,
\eeq
where the only difference is the underlined term. 
The definition of the coordinates $u_1$, $u_2$, $u_3$ and $v_1$ are the same as \eqref{eq:E6incompuiv1}. 
Also, $w$ is the same as \eqref{eq:E6incompw}.
The geometry contains 
a generalized conifold singularity  
${\mathfrak v}_2$ (see the second column of Figure  \ref{Fig:E6incomp}).

%

\bigskip 

\noindent
\underline{Resolution of ${\mathfrak v}_2$
}
\smallskip

The resolution of ${\mathfrak  v}_2$ 
is done in parallel with the incomplete case.
The resulting intersections of $C$'s and $\delta$'s are the same 
as the incomplete case (the third diagram in the lower row of Figure \ref{Fig:E6incomp}).
The limits of $C$'s are also the same as \eqref{eq:E6incomplimit1} and \eqref{eq:E6incomplimit2}.

A difference arises in the property of the remaining singularity ${\mathfrak v}_3$
which resides at the origin $(\xi,u_2,u_3,V_1)=(0,0,0,0)$ in patch $1$ \eqref{eq:E6incompEY1}.
In the incomplete case, ${\mathfrak v}_3$ is a conifold singularity as seen in 
\eqref{eq:E6incompEY1curve} and \eqref{eq:E6incomp1conifold}.
On the other hand, in the complete case, ${\mathfrak v}_3$ is not a conifold singularity, 
since the resolved geometry is given by 
\bes
   \Phi_{zx1}(\xi,u_2,u_3,V_1) 
    & =u_2 u_3 - V_1 \{\xi V_1+(\underline{w^2}+a_{3,2} u_2) \} =0,     \\
 w & = \frac{1}{c}( u_3-\xi-a_{4,3}u_2 -a_{6,5}u_2^2),
\label{eq:E6compEY1curve}
\ees
which is not a conifold: the underlined part $-V_1 w^2$ is now cubic.
Also, it does not look like a generalized conifold singularity form, 
since the cubic terms do not seem to be factorizable.
However, as we will see below, at least at special points of the complex moduli space,
it does have the generalized conifold singularity form and 
the singularity ${\mathfrak v}_3$ is resolved by inserting two $\PP^1$'s.
In other words, ${\mathfrak v}_2$ is an singularity where the three conifold singularities overlap.  

The exceptional sets in this patch are given by 
appropriately replacing $w$ with $w^2$ in the incomplete ones
\eqref{eq:E6incompEY1deltav2}
and 
\eqref{eq:E6incompEY1exceptionalset} (underlined terms):
\bes
&
\begin{cases}
\displaystyle C_1^+\, :\, u_2=0,\, V_1=0, \, w=\frac{1}{c}(u_3-\xi),   \\
\displaystyle C_1^-\, :\, u_2=0,\, \xi V_1=-\underline{w^2}, \, w=\frac{1}{c}(u_3-\xi),
\end{cases}
\hspace{0.5cm} 
  \hspace{0.3cm} 
\delta_1\, : \, u_2=0,\, V_1=0,\, u_3=\xi,   \\
& 
\begin{cases}
\displaystyle C_2^+\, :\, \xi=0,\, u_2 u_3=V_1(\underline{w^2}+a_{3,2}u_2), \, w=\frac{1}{c} ( u_3-a_{4,3}u_2-a_{6,5}u_2^2),  \\
\displaystyle C_2^-\, :\, \mbox{invisible},
\end{cases}
\\
&   
 \hspace{5.3cm}    
\begin{cases}
\delta_2^+ \, : \,  \xi=0,\, u_3 = a_{3,2}V_1,\,  u_3=a_{4,3}u_2+a_{6,5}u_2^2,  \\
\delta_2^- \,: \, \mbox{invisible}.
\end{cases}   \\
&    
\hspace{5.6cm} \,       
\delta_{{\mathfrak v}_2} \,:\, \xi=0,\, u_2 =0,\, u_3=0.
\label{eq:E6compEY1exceptionalset}    
\ees

\bigskip 

\newpage

\noindent
\underline{Resolution of ${\mathfrak v}_3$
}
\smallskip

Let us try to resolve ${\mathfrak v}_3$ by inserting $\PP^1$ in the plane $(u_3,V_1)$ as before.
In patch $1'$, where $(u_3,V_1)=(\eta, \eta V'_1)$, we obtain a regular geometry.  
The intersections and limits are 
the same as those in the incomplete case \eqref{eq:E6incomp11dintersection} and
\eqref{eq:E6incomplimit11d}.
On the other hand, in patch $2'$ with $(u_3,V_1)=(\eta U_3,\eta)$, the resolved geometry 
still contains a codimension-two singularity ${\mathfrak v}_4$.
Replacing $w$ with $w^2$ in \eqref{eq:E6incompzx12dPhi}, we have
\bes
&  \Phi_{zx12'}(\xi,u_2,U_3,\eta) =u_2 U_3 - \{
\xi\eta +(\underline{w^2}+a_{3,2}u_2)\}=0,
\\
&w = \frac{1}{c} (\eta U_3-\xi -a_{4,3}u_2
                                -a_{6,5}u_2^2) . \\
& \mbox{Singularity : }\, {\mathfrak v}_4 = (0,0,a_{3,2},0).
\label{eq:E6compEY11d}
\ees
After the coordinate shift 
\beq
  \widetilde{U}_3 \equiv U_3 - a_{3,2}, 
\eeq
the resolved geometry is given by
\bes
&    \Phi_{zx12'}(\xi,u_2,\widetilde{U}_3,\eta) 
     = u_2\widetilde{U}_3  - \{\xi \eta+w^2 \} =0 ,   \\
& 
w = \frac{1}{c}(\eta \widetilde{U}_3+a_{3,2}\eta-\xi -a_{4,3}u_2
                                -a_{6,5}u_2^2), 
\label{eq:E6compEY11d2}
\ees
where ${\mathfrak v}_4$ is now located at the origin.
In general, it does not have a conifold form because of the subleading terms,
but at a special point of the complex moduli space
\beq
 a_{4,3}=a_{6,5}=0,
\label{eq:E6compamoduli}
\eeq
$\Phi_{zx12'}$ is decomposed 
into the part proportional to $\widetilde{U}_3$ and 
the remaining part of quadratic in $\xi$ and $\eta$, 
and then the geometry is in a conifold form and ${\mathfrak v}_4$ is a conifold singularity.
Explicitly, one can write
\beq
   \Phi_{zx12'} =  \tilde{u}_2 \widetilde{U}_3
                     - \zeta_+ \zeta_- = 0 ,
\label{eq:E6compEY11dconifold}
\eeq
where the coordinates $\tilde{u}_2$ and $\zeta_{\pm}$ are defined by
\bes
  \tilde{u}_2 & \equiv u_2-\frac{1}{c^2} \{ 2\eta (a_{3,2}\eta-\xi) +\eta^2 \widetilde{U}_3 \} , \\
  \zeta_{\pm} & \equiv \frac{1}{c}(\xi +\alpha_{\pm} \eta)
\ees
with
\beq
\alpha_+ + \alpha_-  =  -2a_{3,2}+c^2 \, , \quad \quad
\alpha_+ \alpha_- = a_{3,2}^2.
\label{eq:alphapm}
\eeq
Going back to the geometry $\Phi_{zx1}=0$ \eqref{eq:E6compEY1curve},
it exactly has the generalized conifold form under the specializations 
\eqref{eq:E6compamoduli}, so that 
\beq
   \Phi_{zx1} = \tilde{u}_2 \tilde{u}_3-\frac{1}{c^2} V_1(\xi+\alpha_+ V_1)(\xi+\alpha_- V_1) = 0,
\label{eq:E6compPhizx1EY}
\eeq
where $\tilde{u}_3 = u_3-a_{3,2}V_1$ and $\tilde{u}_2=u_2-\frac{V_1}{c^2} \{\tilde{u}_3+2(a_{3,2}V_1-\xi)\}$.

It should be noted that the specializations \eqref{eq:E6compamoduli} do not change the structure of $E_6$ singularity.
For these specializations, though the order of $f$ enhances to $\infty$ at $w=0$,
the order of $g$ is kept to be $4$ since $a_{3,2}
\neq 0$
(see \eqref{eq:Weierstrass} and Table \ref{singorder}).
Also, all the intersections $C_i\cdot C_j$, $\delta_i\cdot \delta_j$ and the limits 
$\mbox{lim}_{w\rightarrow 0}C_i$ are the same as the generic case\footnote
{Only non-trivial difference is the definition of 
$\delta_2^+$ in patch $2'$ in the second step of
our
resolution \eqref{eq:E6incompEY12dcurve}. It is modified to be 
$\delta_2^+\,:\, \xi=0,\, \eta=0,\, U_3=a_{3,2}$.
Nevertheless, the limit of $C_2^+$ is the same as \eqref{eq:E6incomplimit12d}. }. 
Therefore, nothing changes even if we impose the condition 
\eqref{eq:E6compamoduli} from the beginning.

The 
exceptional sets \eqref{eq:E6compEY1exceptionalset} 
are rewritten in the coordinates $(\widetilde{U}_3, \tilde{u}_2,\zeta_+,\zeta_-)$ as
($C_1^+$, $C_2^-$, $\delta_1$ and $\delta_2^-$ are invisible in this patch)
%
%
%
\bes
& C_1^-\, :\, \tilde{u}_2 =-\frac{1}{c^2} \eta \, \big( 2c \, w -\eta \widetilde{U}_3 \big),
\,  \xi \eta =-w^2, \, w=\frac{1}{c} \{ \eta (\widetilde{U}_3+a_{3,2})-\xi \} ,\\
& C_2^+\, :\, \alpha_+\zeta_-=\alpha_-\zeta_+,\, \widetilde{U}_3 \tilde{u}_2= \zeta_+ \zeta_-,\, 
                  w=\frac{1}{\alpha_+-\alpha_-}(\zeta_+-\zeta_-) (\widetilde{U}_3+a_{3,2}), \\
&   
\delta_2^+ \, : \, \zeta_+=0, \, \zeta_- =0,\,  \widetilde{U}_3=0, \\
& 
\delta_{{\mathfrak v}_2}\,:\, \xi=0,\, 
                                                   \tilde{u}_2=-\frac{16}{c^2}a_{3,2} \eta^2, \, 
                                                   \widetilde{U}_3=-a_{3,2}, \\
& 
\delta_{{\mathfrak v}_3}\,:\, \zeta_+=0,\, \zeta_-=0,\,  \tilde{u}_2=0,
\label{eq:E6compEY12dcurve}
\ees
where $\xi$ and $\eta$ are written by $\zeta_{\pm}$ as
\beq
   \xi=\frac{c}{\alpha_+-\alpha_-}(\alpha_+\zeta_- -\alpha_- \zeta_+),\quad
   \eta=\frac{c}{\alpha_+-\alpha_-}(\zeta_+-\zeta_-).
\label{eq:xieta}
\eeq
In particular, one can show that $\xi=0$ yields $\eta^2 = \frac{c^2}{a_{3,2}^2}\zeta_+\zeta_-$,
which we used  to obtain the form of $C_2^+$ in \eqref{eq:E6compEY12dcurve}.
The intersections and the limits of these objects are the same as those for the incomplete case
\eqref{eq:E6incompint12d} and \eqref{eq:E6incomplimit12d}.
The conifold singularity ${\mathfrak v}_4$ is located at the intersection point of $\delta_2^+$ and $ \delta_{{\mathfrak v}_3}$
as shown in the rightmost diagram in Figure \ref{Fig:E6incomp}.

\bigskip 

\noindent
\underline{Resolution of ${\mathfrak v}_4$}
\smallskip

For the final step of the resolution process, let us resolve the conifold singularity ${\mathfrak v}_4$ 
\eqref{eq:E6compEY11dconifold} by inserting $\PP^1$ in $(\widetilde{U}_3,\zeta_+) $ 
plane as
\beq
   (\widetilde{U}_3,\zeta_+) = (U'_3 \chi, Z_+ \chi).
\eeq
In the coordinate patch $(\widetilde{U}_3,\zeta_+) = (\chi, Z_+ \chi)$  (denoted as patch $1''$),
the resolved geometry is given by
\bes
&  \Phi_{zx12'1''} = \chi^{-1} \Phi_{zx12'} = \tilde{u}_2-Z_+ \zeta_-  = 0.  \\
& \mbox{Singularity~:}\, \mbox{none (regular)},  \\
& C_1^-\, :\, \tilde{u}_2 =-\frac{1}{c^2} \eta \, \big( 2c\, w -\eta \chi \big)    ,
\,  \frac{\xi \eta +w^2}{\chi}=0, \, w=\frac{1}{c} \{ \eta (\chi +a_{3,2})-\xi \} ,\\
& C_2^+\, :\, \alpha_+\zeta_-=\alpha_- Z_+ \chi ,\, \tilde{u}_2= Z_+ \zeta_-,\, 
                  w=\frac{1}{\alpha_+-\alpha_-}(Z_+ \chi -\zeta_-) (\chi+a_{3,2}), \\
&   
\delta_2^+ \, : \, \mbox{invisible} , \\
& 
\delta_{{\mathfrak v}_2}\,:\, \alpha_+ \zeta_- = \alpha_- Z_+ \chi,\, 
                                                   \tilde{u}_2=-\frac{a_{3,2}}{\alpha_-^2} \zeta_-^2, \, 
                                                   \chi=-a_{3,2}, \\
& 
\delta_{{\mathfrak v}_3}\,:\, Z_+=0,\, \zeta_-=0,\,  \tilde{u}_2=0,  \\
& 
\delta_{{\mathfrak v}_4}\,:\, \chi=0,\, \zeta_-=0,\, \tilde{u}_2=0.
\label{eq:E6compEY12d1ddexceptionalset}    
\ees
The intersections are
\beq
    \delta_{{\mathfrak v}_2}\cdot  \delta_{{\mathfrak v}_3}\neq0,\quad   \delta_{{\mathfrak v}_3}\cdot  \delta_{{\mathfrak v}_4} \neq 0.
\label{eq:compE6zx12d1ddint}
\eeq

Let us explain how the second equation of $C_1^-$ is obtained.
In the previous patch, it has the form $\xi \eta +w^2=0$ \eqref{eq:E6compEY12dcurve}.
It is rewritten in the present patch as
\bes
   \xi \eta+w^2 & \sim 
c^2(\alpha_+ \zeta_- -\alpha_- Z_+ \chi)(Z_+\chi-\zeta_-)
                      +\big\{ (Z_+ \chi -\zeta_-)(\chi+a_{3,2})
                                                                  -(\alpha_+ \zeta_- -\alpha_- Z_+ \chi) \big\}^2  
\\
& = \zeta_-^2 \big\{ -c^2 \alpha_+ +(a_{3,2}+\alpha_+)^2 \big\}  + \chi \{\cdots\}  \\
& = \chi \{\cdots\}.   \nonumber
\ees
Here the coefficient of $\zeta_-^2$ vanishes because of \eqref{eq:alphapm}.
Thus 
$C_1^-$ contains the component with $\chi=0$.
It is easily seen that this component is not complex one dimensional but merely a point.
In order to subtract this irrelevant component, the second equation is divided by $\chi$.

Let us evaluate the limit $w \rightarrow 0$ of $C_1^-$ in the present patch. 
Firstly, we estimate the third equation of $C_1^-$ \eqref{eq:E6compEY12d1ddexceptionalset} at $w=0$.
By substituting \eqref{eq:xieta} and $(\widetilde{U}_3,\zeta_+) = (\chi, Z_+ \chi)$ into the equation, 
we find that $\zeta_-$ is factorized by $\chi$ as
\beq
   (\alpha_+ + a_{3,2}) \zeta_- =  \chi \{ Z_+ (\alpha_- +a_{3,2}+\chi) -\zeta_- \} .
\label{eq:C1m3}
\eeq
Then, $\xi$ \eqref{eq:xieta} is also factorized by $\chi$ such that 
\bes
 (\alpha_+ + a_{3,2}) \xi &\sim (\alpha_+ +a_{3,2})(\alpha_+\zeta_- - \alpha_- Z_+ \chi)  \\
                                  &=  \chi \big[ \alpha_+ \{ Z_+ (\alpha_- +a_{3,2}+\chi) -\zeta_- \} 
                                        -(\alpha_+ +a_{3,2}) \alpha_- Z_+   \big]  \\
                                  &=  \chi \alpha_+ \big\{ Z_+ (a_{3,2}\frac{\alpha_+-\alpha_-}{\alpha_+}+\chi) -\zeta_- 
                                          \big\} . 
\ees
Thus the second equation of $C_1^-$ is estimated at $w=0$ as
\beq
   \frac{\xi \eta}{\chi} 
           \sim \big\{ Z_+ (a_{3,2}\frac{\alpha_+-\alpha_-}{\alpha_+}+\chi) -\zeta_- \big\} (Z_+ \chi -\zeta_-) = 0.
\eeq
Plugging back one of its solution $\zeta_- = Z_+ (a_{3,2}\frac{\alpha_+-\alpha_-}{\alpha_+}+\chi)$
into the third equation \eqref{eq:C1m3}, we obtain a simple form
\beq
Z_+(\chi+a_{3,2})=0 .
\eeq
Also, substituting the other solution $\zeta_- =Z_+ \chi$ into \eqref{eq:C1m3}, we have
\beq
  Z_+ \chi = 0.
\eeq
Therefore, 
\bes
 \mbox{lim}_{w\rightarrow 0} \,C_1^- &= \{ \tilde{u}_2 = \frac{1}{c^2} \eta^2 \chi ,\, 
                             \zeta_-=Z_+(a_{3,2}\frac{\alpha_+-\alpha_-}{\alpha_+}+\chi),\, Z_+(\chi+a_{3,2})=0\}  \\
            & \quad  \cup \{ \tilde{u}_2 = \frac{1}{c^2} \eta^2 \chi ,\, 
                             \zeta_-=Z_+ \chi,\, Z_+\chi=0 \}  \\
           & = \{ Z_+=0,\, \zeta_-=0,\, \tilde{u}_2=0 \} 
               \cup \{\chi=-a_{3,2},\, \zeta_-=-Z_+a_{3,2}\frac{\alpha_-}{\alpha_+},\, 
                                                 \tilde{u}_2 = -\frac{1}{c^2}\eta^2 a_{3,2}\}  \\
      & \quad \cup \{ Z_+=0,\, \zeta_-=0,\, \tilde{u}_2=0\}\cup \{\chi=0,\, \zeta_-=0,\, \tilde{u}_2=0\}   \\
          & =  \delta_{{\mathfrak v}_2} \cup 2 \delta_{{\mathfrak v}_3} \cup  \delta_{{\mathfrak v}_4}.
\ees
The limit of $C_2^+$ \eqref{eq:E6compEY12d1ddexceptionalset} is easily calculated as 
\bes
  \mbox{lim}_{w\rightarrow 0} \, C_2^+ &=\{ \alpha_+\zeta_-=\alpha_- Z_+ \chi ,\, \tilde{u}_2= Z_+ \zeta_-,\, 
                       \zeta_- =Z_+ \chi  \}  \\
                       &\quad \cup \{\alpha_+\zeta_-=\alpha_- Z_+ \chi ,\, \tilde{u}_2= Z_+ \zeta_-,\, 
                           \chi=-a_{3,2}\}  \\
                 & = \{Z_+\chi=0,\,\zeta_-=0,\, \tilde{u}_2=0 \}  
            \cup \{ \alpha_+\zeta_-=\alpha_- Z_+ \chi,\, \tilde{u}_2=-\frac{\alpha_+}{a_{3,2}\alpha_-}\zeta_-^2, \,
                        \chi=-a_{3,2}\} \\
                 & = \delta_{{\mathfrak v}_2} \cup  \delta_{{\mathfrak v}_3} \cup  \delta_{{\mathfrak v}_4}.
\ees
Compared with the incomplete case \eqref{eq:E6incomplimit},
one can see that both of the limit of $C_1^-$ and $C_2^+$ are modified to contain  $ \delta_{{\mathfrak v}_4}$.

Similar calculations can be done in the other coordinate patch $(\widetilde{U}_3,\zeta_+)=(U_3' \chi, \chi)$.
In this patch, one can show that 
\beq
\delta_2^+\cdot  \delta_{{\mathfrak v}_4} \neq0.
\label{eq:E6compzx12d2ddint}
\eeq

\begin{figure}[htbp]
  \begin{center}          
         \includegraphics[clip, width=6.6cm]{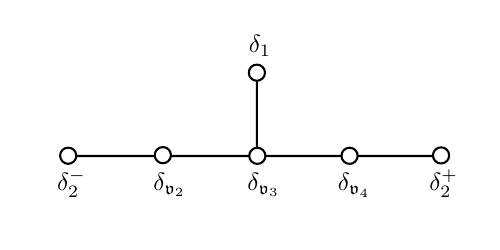}
       \caption{The intersection matrix of $\delta$'s for the complete $E_6$. }
    \label{Fig:E6compCartan}
  \end{center}
\end{figure}

$C_1^+$ and $C_2^-$ are invisible in the both patches and the limits are not modified.
Thus the final result for the complete $E_6$ case is given by
\bes
 C_1^+ & 
 =
 \delta_1 , \\
 C_1^- & =
   \delta_1 +  \delta_{{\mathfrak v}_2} + 2 \delta_{{\mathfrak v}_3}+ \delta_{{\mathfrak v}_4}, \\
 C_2^+ & =
   \delta_2^+ +  \delta_{{\mathfrak v}_2} +  \delta_{{\mathfrak v}_3}+ \delta_{{\mathfrak v}_4}, \\
 C_2^- & =
   \delta_2^-.  
\label{eq:E6complimit}
\ees
One can see from \eqref{eq:compE6zx12d1ddint} \eqref{eq:E6compzx12d2ddint} that $ \delta_{{\mathfrak v}_4}$ is nothing but the missing node 
of the incomplete $E_6$ diagram (Figure \ref{Fig:E6incompCartan}),
and the intersecting pattern of $\delta$'s of the complete $E_6$ is the full $E_6$ Dynkin diagram.
Furthermore, their intersection matrix is identical to the Cartan matrix of $E_6$.
Namely, under the identifications \eqref{eq:E6complimit}, the $SU(5)$ Cartan matrix for $C$'s is reproduced 
if the intersection matrix of $\delta$'s is the ordinary Cartan matrix of $E_6$ as depicted in Figure \ref{Fig:E6compCartan}.

Since $\delta$'s form the full $E_6$ Dynkin diagram, the localized matter is the full coset \eqref{eq:E6cosetspec}.
This is also equivalent to the spectrum expected from the anomaly-free condition (see 
Table \ref{E6pattern}). 

%

\subsection{$SU(5) \rightarrow E_7$}
\label{sec:E7}

\subsubsection{Incomplete and complete $E_7$ singularities}
\label{section:incompleteandcompleteE7}
As we have seen in the previous section, the only relevant patch after the two-time
$\PP^1$ insertions (small resolutions) is the one with $U_1=V'_1=1$,
in which the proper transform is given by (\ref{Phizxxieta}) with (\ref{xiconstraint}).
After all, the difference of the singularities only arises through the orders of various 
sections in $w$. In the previous $E_6$ examples, if we assume $\mbox{o}_w(a_{1,0})=1$,
$\mbox{o}_w(a_{2,1})=1$ and other sections $a_{3,2}$, $a_{4,3}$, $a_{6,5}$ are nonzero 
at $w=0$, we realize an incomplete $E_6$ singularity, while if we consider  
$\mbox{o}_w(a_{1,0})=2$ and $\mbox{o}_w(a_{2,1})=1$ with the same other $a_{i,j}$'s 
we get a complete $E_6$ singularity. 

In fact, what kind of singularities remain after the two small resolutions is 
fairly obvious from  (\ref{Phizxxieta}) and (\ref{xiconstraint}). Indeed, since $\Phi_{zx\xi\eta}$  
(\ref{Phizxxieta}) contains the term $-a_{1,0}(w)$, 
one can immediately see that $\Phi_{zx\xi\eta}=0$ is regular if $\mbox{o}_w(a_{1,0})=1$, 
so no additional singularity arises in the incomplete $E_6$ case. 
Also, if $\mbox{o}_w(a_{1,0})=2$, the lowest-order terms in $(U_3,\eta,z_2,w)$ are 
quadratic, so one sees that there remains a conifold singularity in the complete $E_6$ case. 
So  let us first examine what properties the incomplete and complete $E_7$ singularities have using (\ref{Phizxxieta}) and (\ref{xiconstraint}) before analyzing their structures in detail.

As we discussed in section \ref{section:incompleteE6E7E8}, 
the ``mildest'' incomplete $E_7$ singularity, the incomplete $1$ singularity, 
can be achieved by assuming the orders of the sections to be
$(\mbox{o}_w(a_{1,0}),\mbox{o}_w(a_{2,1}),\\ 
\mbox{o}_w(a_{3,2}),\mbox{o}_w(a_{4,3}),\mbox{o}_w(a_{6,5})=(1,1,1,0,0)$ (Table \ref{E7pattern}).
This is just an incomplete $E_6$ singularity with $\mbox{o}_w(a_{3,2})$ set to $1$,
in particular $\mbox{o}_w(a_{1,0})$ is $1$. Therefore, the threefold is still regular 
after the two-time small resolutions. We will see, however, that the intersection matrix 
changes. 

The other incomplete $E_7$ singularities, the incomplete $2$, $3$ and $4$ singularities, 
all require that the order of $a_{2,1}$ be $1$. This means that $\xi$ is of order $1$ in $w$. 
Then $\Phi_{zx\xi\eta}$ includes terms like $z_2U_3 - \eta\xi$, so they all gives a conifold 
singularity after the two-time small resolutions.

Finally, a complete $E_7$ singularity is realized by taking the order of $a_{2,1}$ to be $2$ 
in $w$. In this case (\ref{xiconstraint}) indicates that $\xi$ is quadratic in $U_3,\eta,z_2,w$, 
therefore $\xi\eta$ in $\Phi_{zx\xi\eta}$ is cubic in $U_3,\eta,z_2,w$. This is neither a 
conifold singularity nor a generalized conifold singularity in general.

\subsubsection{Incomplete $1$ $E_7$}
\label{sec:incompE7}

The incomplete $1$ $E_7$ geometry 
can be obtained by setting $a_{1,0}=w$ 
and
\bes
  a_{2,1} & = w \, c
  , \\
  a_{3,2} & = w \, d
  .
\ees
The equation 
(\ref{Phi}) of this geometry is given by 
\beq
  \Phi \equiv -(y^2 + w x y +  w \, d \,z^2 y ) +x^3 +w \,c \,z x^2+a_{4,3} z^3 x +a_{6,5} z^5 = 0.
\label{eq:E7incompPhi}
\eeq
It has a codimension-one singularity ${\mathfrak p}_0$ at $(x,y,z,w)=(0,0,0,w)$.

\begin{figure}[htbp]
  \begin{center}          
         \includegraphics[clip, width=15.6cm]{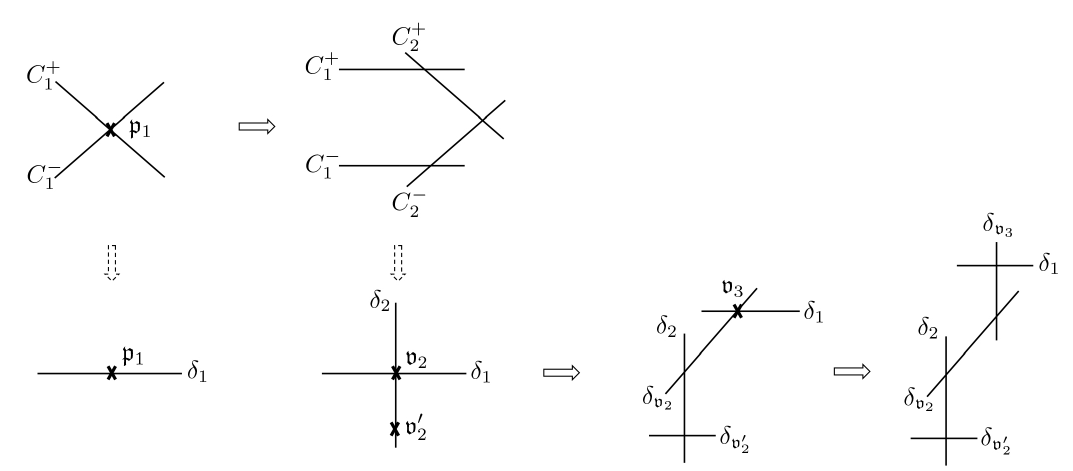}
       \caption{The resolution process of the incomplete 1 $E_7$.} 
    \label{Fig:E7incomp}
  \end{center}
\end{figure}

Since \eqref{eq:E7incompPhi} is obtained simply by a substitution 
$a_{3,2}=w \, d$ in $\Phi$ 
of the incomplete $E_6$, the same substitution 
in the resolved incomplete $E_6$ geometry
basically gives the resolved geometry for this incomplete 1 $E_7$ singularity.
Below, we focus on the difference from the incomplete $E_6$ case.
The whole process of the resolution is depicted in Figure \ref{Fig:E7incomp}.

\bigskip

\noindent
\underline{Blow up of ${\mathfrak p}_0$}
\smallskip

The geometry after blowing up ${\mathfrak p}_0$ is given by
\bes
\Phi_z & \equiv z^{-2} \Phi \\
          & =-y_1\{y_1+ w (x_1 + d \, z)\} +z(x_1^3+w \, c \, x_1^2+a_{4,3} z x_1 +a_{6,5}z^2) =0.
\ees
There is a codimension-one singularity ${\mathfrak p}_1=(0,0,0,w)$, which we blow up next.

\bigskip

\noindent
\underline{Blow up of ${\mathfrak p}_1$}
\smallskip

After the blow up of  ${\mathfrak p}_1$, two differences arise.
The first difference is that two exceptional sets $C_2^{\pm}$ 
are combined into one exceptional 
curve $\delta_2$ at $w=0$; they remained split as $\delta_2^{\pm}$ in the incomplete $E_6$ case.
The other difference is that a new codimension-two singularity ${\mathfrak v}_2'$ arises 
(see the second column of Figure \ref{Fig:E7incomp}).  
 
In the patch $(x_1,y_2,z_2)$ with $(x_1,y_1,z)=(x_1, y_2 x_1, z_2 x_1)$, 
the geometric data are given by substituting $a_{3,2} = w\, d$ into \eqref{eq:E6incompzxCdelta}.
Then one can easily see that $\delta_2^{\pm}$ become a single $\delta_2$.
There are two codimension-two singularities 
\beq
{\mathfrak v}_2 \,:\, (x_1,y_2,z_2,w) = (0,0,0,0), \quad \quad {\mathfrak v}_2'\,:\, (0,0,-\frac{a_{4,3}}{a_{6,5}},0).
\label{eq:v2v2d}
\eeq
${\mathfrak v}_2'$ is an ordinary conifold singularity (see \eqref{eq:E7incompzz} below), whereas ${\mathfrak v}_2$ is 
a generalized conifold 
singularity
since the geometry has the form similar to \eqref{eq:E6incompzx} :
\beq
   \Phi_{zx}(u_1,u_2,u_3,v_1)= u_1u_2u_3-v_1\{v_1+w(1+d\,u_2)\}.
\label{eq:E7incompzx}
\eeq
Here the definitions of $u_i$ and $v_i$ are the same as \eqref{eq:E6incompuiv1} and \eqref{eq:E6incompw}. 

In the other patch $({x}_2,{y}_2,z)$ with $(x_1,y_1,z)=({x}_2 z, {y}_2 z, z)$, 
it is
easily seen that ${\mathfrak v}_2'$ is a conifold singularity:
\bes
& \Phi_{zz} ({x}_2,{y}_2,z,w)\equiv z^{-2} \Phi_z({x}_2 z, {y}_2 z, z,w) \\
& \hspace{2.7cm}           =z(a_{6,5}+a_{4,3}{x}_2+{x}_2^3 z+w\,c\,{x}_2^2) 
                                    -{y}_2\{{y}_2+w({x}_2+d)\} =0 . \\
& \mbox{Singularity} \,: \,  {\mathfrak v}_2' = (-\frac{a_{6,5}}{a_{4,3}},0,0,0) ,\\
& 
\begin{cases}
C_2^{+}\,:\, z=0 ,\,\, {y}_2 = 0 \\
C_2^{-}\,:\, z=0, \,\, {y}_2 = -w ({x}_2+d) 
\end{cases}
    \rightarrow \quad \delta_2 \, :\, z=0,\, {y}_2 = 0,\, w=0 , \\
& \hspace{0.4cm} C_1^{\pm}\,:\, \mbox{invisible},    \hspace{3.9cm}    \delta_1\, :\, \mbox{invisible}.
\label{eq:E7incompzz}
\ees

\bigskip

\noindent
\underline{Small resolution of ${\mathfrak v}_2'$}
\smallskip

The conifold singularity ${\mathfrak v}_2'$ is removed by the standard small resolution.
The geometry  \eqref{eq:E7incompzz} is written as
\beq
\Phi_{zz} = s_1s_2-t_1t_2 = 0
\eeq
by using the coordinates
\bes
   & s_1 \equiv z, \\
   & s_2 \equiv a_{4,3} ({x}_2+\frac{a_{6,5}}{a_{4,3}})+{x}_2^3z+w\,c\,x_2^2 , \\
   & t_1 \equiv y_2 , \\
   & t_2 \equiv y_2+w(x_2+d).
\ees
${\mathfrak v}_2'$ is located at the origin.
In these coordinates, 
\bes
&
\begin{cases}
 C_2^+~:~ s_1=0, \, t_1 =0, \, t_2 \propto w \\
 C_2^-~:~ s_1=0, \, t_2 = 0,\, t_1 \propto -w ,
 \end{cases}
 \hspace{0.5cm} 
 \rightarrow    \hspace{0.2cm}   
  \delta_2~:~ s_1=0,\, t_1=0,\, t_2=0.  
\ees
Choosing the $(s_2,t_2)$ plane for the insertion of $\PP^1$ (denoted as $\delta_{{\mathfrak v}'_2}$)
and evaluating the 
expressions 
of $C_2^{\pm}$ and $\delta_2$ into the resolved space, 
one can easily find that 
\beq
    \delta_2\cdot \delta_{{\mathfrak v}_2'} \neq 0,
\eeq
and 
\bes
   &C_2^+ \rightarrow  \delta_2 + \delta_{{\mathfrak v}_2'},  \\
   &C_2^- \rightarrow \delta_2.
\label{eq:E7incomplimit1}
\ees

\bigskip

\noindent
\underline{Resolution of ${\mathfrak v}_2$}
\smallskip

The 
resolution of ${\mathfrak v}_2$ proceeds similarly to the incomplete $E_6$ case. 
In patch 1 with $(u_1,v_1)=(\xi , \xi V_1)$, 
the geometry after the resolution of \eqref{eq:E7incompzx} has the form similar to \eqref{eq:E6incompEY1curve} as
\bes
 & \Phi_{zx1}(\xi,u_2,u_3,V_1)  = u_2 u_3-V_1\{ \xi V_1 +w(1+d\, u_2)\} =0,  \\
 &    \hspace{2.6cm}         w  = \frac{1}{c} (u_3-\xi-a_{4,3}u_2-a_{6,5}u_2^2) .\\
 & \mbox{Singularity~:~} {\mathfrak v}_3=(0,0,0,0) .
 \label{eq:E7incompcurvezx1}
\ees
The exceptional curve $ \delta_{{\mathfrak v}_2}$ is the same as \eqref{eq:E6incompEY1deltav2}
\beq
   \delta_{{\mathfrak v}_2}\,:\, \xi=0,\,u_2=0,\,u_3=0.
\label{eq:E7incompdeltav2}
\eeq
Also,  $\delta_{{\mathfrak v}_2'}$ is translated into this coordinate patch as\,
\beq
   \delta_{{\mathfrak v}_2'}\,:\, \xi=0,\, u_2 = -\frac{a_{4,3}}{a_{6,5}},\, u_3=0.
\eeq
%
$C_1^{\pm}$ and $\delta_1$ are the same as those in \eqref{eq:E6incompEY1exceptionalset},
while 
$C_2^+$ and $\delta_2$ are modified to be 
\bes
 C_2^+\,:\, \xi=0,\, u_2u_3=V_1w(1+d\, u_2),\, w=\frac{1}{c}(u_3-a_{4,3}u_2-a_{6,5}u_2^2),\quad
 \delta_2\,:\, \mbox{invisible}.
\label{eq:E7incompdelta2}
\ees  
${\mathfrak v}_3$ is located at the intersection point of $ \delta_{{\mathfrak v}_2}$ and $\delta_1$:
\beq
     \delta_{{\mathfrak v}_2} \cdot \delta_1 \neq 0.
\eeq
In contrast to the incomplete $E_6$ case, the right hand side of the second equation of $C_2^+$ 
is proportional to $w$ (since $a_{3,2}$ is replaced by $w\,d$) and drops for $w\rightarrow 0$. 
Thus the limit of $C_2^+$ differs from the incomplete $E_6$ case \eqref{eq:E6incompC2plimit} as 
\bes
 \mbox{lim}_{w\rightarrow 0} \, C_2^+ &= \{\xi=0,\, u_2 u_3 = 0,\, u_3=a_{4,3}u_2+a_{6,5}u_2^2 \} \\
                                               &= \{ \xi=0,\, u_2^2(a_{4,3}+a_{6,5}u_2)=0,\, u_3= u_2 (a_{4,3}+a_{6,5}u_2)\}  \\
                                          &= \{\xi=0,\,u_2=0,\, u_3=0\}^{\otimes 2}  \cup \{\xi=0,\, u_2 = -\frac{a_{4,3}}{a_{6,5}},\, u_3=0\} \\
                                               &= 2  \delta_{{\mathfrak v}_2} \cup \delta_{{\mathfrak v}_2'}.
\ees
Namely,
\beq
 C_2^+ \rightarrow 2  \delta_{{\mathfrak v}_2} + \delta_{{\mathfrak v}_2'},
 \label{eq:E7incomplimit2}
\eeq
where the degeneracy of $ \delta_{{\mathfrak v}_2}$ is two, which was one for the incomplete $E_6$ case 
\eqref{eq:E6incomplimit1}.
In patch 2, we obtain the similar modifications. 

\bigskip

\noindent
\underline{Resolution of ${\mathfrak v}_3$}
\smallskip

As we can see from \eqref{eq:E7incompcurvezx1},  ${\mathfrak v}_3$ is a conifold singularity
and is removed by a standard small resolution.
In patch $1'$, the exceptional curve $ \delta_{{\mathfrak v}_3}$ is the same as the one in \eqref{eq:E6incompzx11d}:
\beq
    \delta_{{\mathfrak v}_3}\,:\, \xi=0,\, \eta=0,\, u_2=0.
\eeq
Substituting $a_{3,2}=w \,d$ into the 
incomplete $E_6$ results \eqref{eq:E6incompzx11d}, we find 
\bes
 C_2^+\,:\, \xi=0,\, u_2=V_1'w(1+d u_2),\, w=\frac{1}{c}(\eta-a_{4,3}u_2-a_{6,5}u_2^2),\quad
 \delta_2\,:\, \mbox{invisible},
\label{eq:E7incompdelta21}
\ees  
yielding
\beq
 C_2^+ \rightarrow  \delta_{{\mathfrak v}_3}.
\eeq
The limits of $C_1^{\pm}$ are the same as \eqref{eq:E6incomplimit11d}.
The other patch $2'$ gives the similar result.

\begin{figure}[htbp]
  \begin{center}          
      \vspace{0.7cm}   \includegraphics[clip, width=8.6cm]{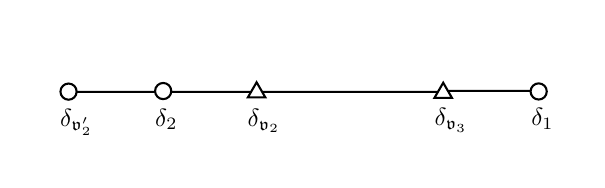}
       \caption{The intersection matrix of $\delta$'s for the incomplete 1 $E_7$. The triangular nodes 
                    have self-intersection numbers $-\frac{4}{3}$ and their mutual intersection numbers are 
                    $\frac{2}{3}$. }
    \label{Fig:E7incompCartan}
  \end{center}
\end{figure}

This completes the resolution of the incomplete 1 $E_7$.
The differences from the incomplete $E_6$ result \eqref{eq:E6incomplimit} are the limits of
$C_2^{+}$ in \eqref{eq:E7incomplimit1} and \eqref{eq:E7incomplimit2}. 
The final forms are 
\bes
 & C_1^+ =
 \delta_1,  \\
 & C_1^- =
 \delta_1 +  \delta_{{\mathfrak v}_2} +2 \delta_{{\mathfrak v}_3}, \\
 & C_2^+ =
 \delta_2 + 2  \delta_{{\mathfrak v}_2} + \delta_{{\mathfrak v}_3} + \delta_{{\mathfrak v}_2'} , \\
 & C_2^- =
 \delta_2 . 
\ees
The intersecting pattern of $\delta$'s forms the $A_5$ Dynkin diagram as depicted in the rightmost 
diagram of Figure \ref{Fig:E7incomp}.
If its intersection matrix has the form
\beq
   \delta_i\cdot \delta_j 
 = - \left( 
\begin{array}{rrrrr}
  2    &      -1        &         0        &    0              &    0     \\
 -1   &       2         &       -1        &    0              &    0     \\
  0    &     -1         & \frac{4}{3}    &  -\frac{2}{3}  &    0     \\
  0    &       0         & -\frac{2}{3}  &    \frac{4}{3}  &     -1                \\
  0    &       0        &         0        &    -1             &     2     
\end{array} 
    \right),
\label{eq:E7incompdeltaCartan}
\eeq
where rows and columns are ordered as $\delta_{{\mathfrak v}_2'}, \delta_{2},  \delta_{{\mathfrak v}_2},  \delta_{{\mathfrak v}_3}$ and $\delta_{1}$,
one can easily check that $C_i$'s form (the minus of) the 
$SU(5)$ Cartan matrix.
The intersections of $\delta$'s are depicted in Figure \ref{Fig:E7incompCartan}.
Two nodes of the $E_7$ Dynkin diagram (the branching out node and its joint node)
are removed.

As before, charged matter spectrum is identified with the set of curves $J$ \eqref{eq:J} 
with $-2 \leq J.J < 0$.
It is shown that possible such self-intersection numbers are $-2$ and $-\frac{4}{3}$;  
the numbers of such curves are $30$ and $30$, respectively.
Again, 
the former curves are ``adjoint of incomplete 1 $E_7$'' and taking their coset 
by $SU(5)$ gives $\bold{5}$, 
while the latter form $\bold{10}\oplus {\bf 5}$: 
\bes
  & \sharp(J.J= -2) = 30  \rightarrow  \bold{5},  \\
  & \sharp(J.J=-\frac{4}{3})  = 30 \rightarrow  \bold{10}\oplus {\bf 5}.
\ees
In all, an incomplete 1 $E_7$ singularity gives matter multiplet $\bold{10} \oplus 2\cdot \bold{5}$,
which exactly coincide with the expected result from the anomaly-free condition (see Table \ref{E7pattern}).
\subsubsection{Incomplete $2$,\,$3$,\,$4$ and complete $E_7$ singularities}
As we discussed in section \ref{section:incompleteandcompleteE7}, a conifold 
singularity remains after the two small resolutions for the incomplete $2$,\,$3$ and $4$ singularities.
Although one could resolve this conifold singularity by an additional small resolution in principle,
it is not easy to compute the representations of charged matter 
because the subleading (cubic or higher order) terms do not fit in the standard conifold form\footnote{
To obtain the charged matter representations, we need the intersection matrix, which
can be read from how $C_i$' s factorize into $\delta_i$' s for $w\rightarrow$ 0.
As we have seen so far, these calculations had relied on the fact that the geometry had 
taken the standard conifold form. One may wonder if one can carry out the similar
calculations just by truncating the higher order terms, but it does not work,
since the truncation of higher order terms in the defining equations of $C_i$'s and $\delta_i$'s generically 
changes the structure of the factorizations.}.
Still, we can speculate what charged matter is generated 
from these singularities as follows.
The exceptional curves $\delta$'s are all connected in the same way in the incomplete $2$,\,$3$ and $4$.
The number of nodes of their intersection diagrams is $6$, since one node ($\PP^1$) is
added to the incomplete 1 diagram shown in Figure \ref{Fig:E7incompCartan}.
It is likely that such diagram is the one that the branching out node of $E_7$ Dynkin 
diagram is removed.
It is the same diagram as the incomplete $SO(10)$ singularity enhanced to $E_7$.
If so, using the result of our previous paper \cite{halfhyper}, 
we expect to obtain $2\cdot {\bf 10}\oplus 3\cdot {\bf 5}$ for these singularities\footnote{The corresponding diagram is the bottom one of Figure 2 of  \cite{halfhyper}
and the number of the curves $J$ obtained from that diagram is 60 ($J.J=-2$) or 32 ($J.J=-3/2$),
as seen in (3.45) of \cite{halfhyper}. 
Decomposing them into $SU(5)$, we obtain $\bold{10}\oplus 2\cdot {\bf 5}$ from the former 
and $\bold{10}\oplus {\bf 5}$ from the latter.
}.
This result just saturates the required matter content for the incomplete $2$ $E_7$ singularity,
but falls short for the incomplete $3$ and $4$ singularities (see Table \ref{E7pattern}).

Note that this does not imply that the anomaly 
cancellation breaks down. In the present case, several exceptional curves 
overlap to form an identical single curve, but we just don't know how to discuss 
what hypermultiplets result from such geometries, or rather, 
we do not have the logic to explain the mechanism that generates 
the amount of matter necessary for anomaly cancellation from such geometries.

We also saw in section \ref{section:incompleteandcompleteE7} 
 that the complete $E_7$ singularity ends up with a singularity that is 
 neither a conifold singularity nor a generalized conifold singularity. 
Such a singularity cannot be resolved by a small resolution.
Therefore, in this case, the required matter generation cannot be explained 
from the set of exceptional curves that arise there,
and in fact, to resolve this singularity, it would be necessary to insert $\PP^2$, 
which would break supersymmetry. This is because, in this case, 
we cannot factor out the enough divisors necessary to obtain a proper transform 
that preserves the canonical class.

\subsection{$SU(5) \rightarrow E_8$}

\subsubsection{Incomplete and complete $E_8$ singularities}
\label{section:incompleteandcompleteE8}

Similarly to the $E_7$ case, general properties of incomplete and complete 
$E_8$ singularities can be derived from (\ref{Phizxxieta}) and (\ref{xiconstraint}). 
From Table \ref{E8pattern}, we see that an incomplete $1$ $E_8$ singularity can be realized 
by requiring that an incomplete $1$ $E_7$ singularity have $a_{4,3}$ that vanishes at $w=0$,
in particular $\mbox{o}_w(a_{1,0})$ remains $1$. Therefore it is regular after the two-time 
small resolutions.

All other incomplete $E_8$ singularities have $\mbox{o}_w(a_{1,0})\geq 2$,  so 
there remains a singularity after the small resolutions. 
The necessary condition for this to be a conifold singularity is 
that the order of $a_{2,1}$ be $1$, and several patterns satisfy this condition.
Otherwise, the incomplete singularities that do not satisfy this, as well as the complete singularity, 
are neither conifold singularities nor generalized conifold singularities, and therefore cannot be 
resolved by a small resolution. 
\subsubsection{Incomplete $1$ $E_8$}

Let us consider the incomplete $1$ $E_8$ geometry. 
Setting $a_{1,0}=w$ and
\bes
  a_{2,1} & = w \, c,
  \\
  a_{3,2} & = w \, d,
  \\
  a_{4,3} & = w \, e,
\ees
the 
equation of this geometry is given by 
\beq
  \Phi \equiv -(y^2 + w x y +  w \, d \,z^2 y ) +x^3 +w \,c \,z x^2+ w \,e \,z^3 x +a_{6,5} z^5 = 0.
\label{eq:E8incompPhi}
\eeq

\begin{figure}[htbp]
  \begin{center}          
         \includegraphics[clip, width=15.6cm]{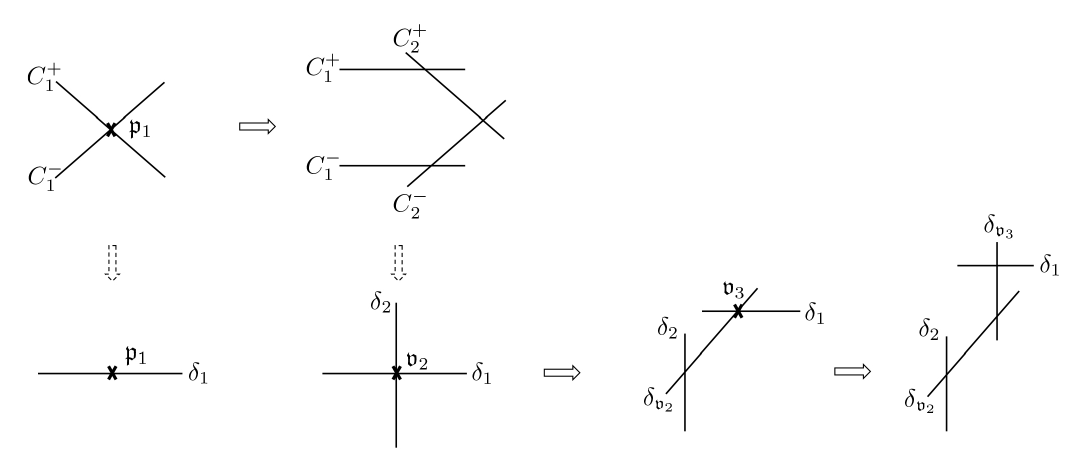}
       \caption{The resolution process of the incomplete 1 $E_8$.} 
    \label{Fig:E8incomp}
  \end{center}
\end{figure}

The resolution proceeds similarly to the incomplete 1 $E_7$ case (section \ref{sec:incompE7})
and the whole process of the resolution is depicted in Figure \ref{Fig:E8incomp}.
The first difference arises after blowing up ${\mathfrak p}_1$.

\bigskip


\noindent
\underline{Blow up of ${\mathfrak p}_1$}
\smallskip

In this case, the codimension-two singularity ${\mathfrak v}_2'$ existed in the incomplete 1 $E_7$ case coincides with ${\mathfrak v}_2$
(see \eqref{eq:v2v2d}).
The geometry is given by
\bes
& \Phi_{zx} =  x_1 z_2 (x_1 + c \, w + e \,w \,z_2 +a_{6,5} \, z_2^2 ) 
                  - y_2 \big\{ y_2+w (1+d \, z_2)\big\}
                       = 0.  \\
& \mbox{Singularity} \,: \, {\mathfrak v}_2 = (0,0,0,0) .       
\label{eq:E8incompzx}
\ees
The exceptional sets are given by \eqref{eq:E6incompzxCdelta} with $a_{3,2} = w\, d$
(see the second column of Figure \ref{Fig:E8incomp}).

\bigskip

\noindent
\underline{Resolution of ${\mathfrak v}_2$}
\smallskip

Let us insert a $\PP^1$ ($= \delta_{{\mathfrak v}_2}$) in the $(x_1,y_2)$ plane as
\beq
  (x_1,y_2) = (\xi U_1, \xi V_1).
\eeq
In the patch $(x_1,y_2)=(\xi,\xi V_1)$, the resolved geometry is given by
\bes
  & \Phi_{zx1}(\xi,z_2,V_1,w)  = z_2 \big\{ \xi +c\, w+ e\, w \,z_2+a_{6,5}z_2^2\big\}
               -V_1\{\xi V_1+w(1+d\, z_2)\}=0.  \\
  & \mbox{Singularity~:~} {\mathfrak v}_3 = (0,0,0,0).  
\ees
Now let us switch variables from $(z_2,\xi,V_1,w)$ to $(u_2,u_3,V_1,w)$ with
\bes
   u_2 & = z_2,  \\
   u_3 & = \xi  +c\, w+ e\, w \,z_2+a_{6,5}z_2^2
\ees
and write the geometry as 
\bes
 \Phi_{zx1}(u_3 ,u_2,V_1,w) & = u_2 u_3 -V_1\{\xi V_1+w(1+d\, u_2)\}=0,  \\
    \xi & = u_3-c\, w - e\, w \,u_2 - a_{6,5}u_2^2.   
\label{eq:E8incompzx1}
\ees
In these variables, the exceptional sets are given by (see \eqref{eq:E6incompEY1exceptionalset})
\bes
& 
\begin{cases}
C_1^+\,:\, u_2=0,\, V_1=0,   \\
C_1^-\,:\, u_2=0,\, (u_3-c w) V_1 = -w,   
\end{cases} 
  \hspace{2.0cm} 
  \delta_1\,:\, u_2=0, \, V_1=0,\, w=0,\\
&
\begin{cases}
C_2^+\,:\,  u_3-c\, w - e\, w \,u_2 - a_{6,5}u_2^2 = 0,\, u_2 u_3=V_1 w(1+d u_2),    \\
C_2^-\,:\, \mbox{invisible},  
\end{cases}
\delta_2\,:\, \mbox{invisible} ,  \\
& \hspace{8.3cm}  \delta_{{\mathfrak v}_2}\,:\, u_2=0,\, u_3=0,\, w=0.  
\label{eq:E8incompzx1sets}
\ees
%
${\mathfrak v}_3$ is located at the intersection point of $\delta_1\cdot  \delta_{{\mathfrak v}_2} \neq0$. 
One can easily see that the limit of $C_1^+$ and $C_2^{\pm}$ are 
\beq
 C_1^+\rightarrow \delta_{1}, \quad 
   C_1^- \rightarrow \delta_1 +  \delta_{{\mathfrak v}_2}, \quad     C_2^+ \rightarrow 3 \delta_{{\mathfrak v}_2}.
\label{eq:E8incomplimit1}
\eeq 
In the patch $(x_1,y_2)=(\xi U_1,\xi )$, there is no singularity and we obtain 
$\delta_2\cdot  \delta_{{\mathfrak v}_2} \neq0$  and  
\beq
   C_1^-\rightarrow  \delta_{{\mathfrak v}_2},\quad C_2^+ \rightarrow \delta_{2}+3 \delta_{{\mathfrak v}_2},\quad  
   C_2^-\rightarrow \delta_2.
\label{eq:E8incomplimit2}
\eeq

\bigskip

\noindent
\underline{Resolution of ${\mathfrak v}_3$}
\smallskip

Since the geometry 
\eqref{eq:E8incompzx1} is in the
conifold form, 
${\mathfrak v}_3$ is removed by the small resolution, which completes the resolution process.
As before, let us insert a $\PP^1$ (= $\delta_{{\mathfrak v}_3}$) in the $(u_3,V_1)$ plane
\beq
  (u_3,V_1) = (\eta U_3, \eta V'_1).
\eeq 
In the patch $(u_3,V_1) = (\eta, \eta V'_1)$, 
the exceptional sets are given by
\bes
& 
\begin{cases}
  C_1^+\,:\, u_2=0,\, V'_1=0, \\
  C_1^-\,:\, u_2=0,\, (\eta-c w)\eta V'_1=-w,
\end{cases} 
\hspace{1.9cm} 
\delta_1\,:\, u_2=0,\, V'_1=0,\, w=0,  \\
& \hspace{0.4cm}C_2^+\,:\,  \eta-c\, w - e\, w \,u_2 - a_{6,5}u_2^2 = 0,\, u_2 =V'_1 w(1+d u_2), \\
& \hspace{8.3cm}  \delta_{{\mathfrak v}_2}\,:\, \mbox{invisible},   \\
  &    \hspace{8.3cm}  \delta_{{\mathfrak v}_3}\,:\, u_2=0,\,\eta=0,\, w=0.    
\ees
$ \delta_{{\mathfrak v}_3}$ intersects with $\delta_1$ and the limits are given by 
\beq
  C_1^+\rightarrow \delta_1, \quad  C_1^- \rightarrow \delta_1 +2  \delta_{{\mathfrak v}_3} , \quad C_2^+ \rightarrow  \delta_{{\mathfrak v}_3}.
\label{eq:E8incomplimit3}
\eeq
Similarly, in another patch 
$(u_3,V_1) = (\eta U_3,\eta)$,
$ \delta_{{\mathfrak v}_3}$ intersects with $ \delta_{{\mathfrak v}_2}$ and the limits are ($C_1^+$ is invisible)
\beq
  C_1^-\rightarrow  \delta_{{\mathfrak v}_2}+2 \delta_{{\mathfrak v}_3}, \quad C_2^+ \rightarrow 3 \delta_{{\mathfrak v}_2}+ \delta_{{\mathfrak v}_3}.
\label{eq:E8incomplimit4}
\eeq

\begin{figure}[htbp]
  \begin{center}          
      \vspace{0.7cm}   \includegraphics[clip, width=7.2cm]{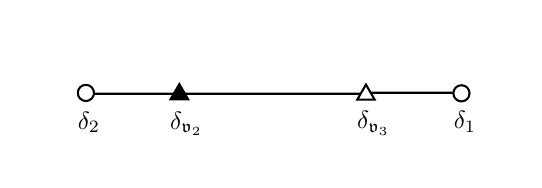}
       \caption{The intersection matrix of $\delta$'s for the incomplete 1 $E_8$. Triangular nodes have 
                   self-intersection number $-\frac{4}{5}$ (black) and $-\frac{6}{5}$ (white). Their mutual intersection number 
                   is $\frac{2}{5}$.}
    \label{Fig:E8incompCartan}
  \end{center}
\end{figure}

By taking the union of \eqref{eq:E8incomplimit1}, \eqref{eq:E8incomplimit2}, \eqref{eq:E8incomplimit3} and 
\eqref{eq:E8incomplimit4}, we obtain the final result for the limits of $C$'s as
\bes
 & C_1^+=
  \delta_1,  \\
 & C_1^-=
  \delta_1+ \delta_{{\mathfrak v}_2}+2 \delta_{{\mathfrak v}_3},  \\
 & C_2^+=
  \delta_2+3 \delta_{{\mathfrak v}_2}+ \delta_{{\mathfrak v}_3}, \\
 & C_2^-=
  \delta_2.
\ees 
The intersecting pattern of the four $\delta$'s is depicted in the rightmost diagram of Figure \ref{Fig:E8incomp}.
It is easily shown that the intersections of $C$'s form (minus of) the $SU(5)$ Cartan 
matrix iff $\delta$'s have the intersection matrix
\beq
   \delta_i\cdot \delta_j 
 = - \left( 
\begin{array}{rrrr}
   2         &       -1        &    0              &    0     \\
  -1        & \frac{4}{5}    &  -\frac{2}{5}  &      0     \\
   0         & -\frac{2}{5}  &    \frac{6}{5}  &      -1               \\
   0        &       0           &      -1           &      2     
\end{array} 
    \right),
\label{eq:E8incompdeltaCartan}
\eeq
where rows and columns are ordered as $\delta_{2},  \delta_{{\mathfrak v}_2},  \delta_{{\mathfrak v}_3}$ and $\delta_{1}$.
The result is summarized in Figure \ref{Fig:E8incompCartan}.

One can show that possible self-intersections of the curves $J$ \eqref{eq:J} 
with $-2 \leq J.J < 0$
are $-2$, $-\frac{6}{5}$ and $-\frac{4}{5}$;  
the numbers of such curves are $20$, $20$ and $10$, respectively.
From these curves, we expect to obtain 
the following charged representations:
\bes
  & \sharp(J.J= -2) = 20  \rightarrow  \mbox{none},  \\
  & \sharp(J.J=-\frac{6}{5})  = 20 \rightarrow  \bold{10},\\
  & \sharp(J.J=-\frac{4}{5})  = 10 \rightarrow  \bold{5}.\\
\ees
That is, an incomplete 1 $E_8$ singularity gives matter multiplets $\bold{10} \oplus \bold{5}$,
which are, again,  
{\it less than} the expected result from the anomaly-free condition; 
${\bf 5}$ is missing here (see Table \ref{E8pattern}).


It may seem counterintuitive that as the singularity worsens, 
the number of exceptional curves decreases;
although ${\mathfrak v}_2'$ that existed in the incomplete $1$ $E_7$ 
overlaps with ${\mathfrak v}_2$ in the incomplete $1$ $E_8$
, the subsequent small 
resolution can be done in a same way.

\subsubsection{Other incomplete  and complete $E_8$ singularities}
We have discussed in \ref{section:incompleteandcompleteE8} that 
some of the incomplete $E_8$ singularities leave a conifold singularity 
after the two small resolutions. 
In this case, even if it could be resolved by a further small resolution, 
it would at best add one more exceptional curve, which is generically insufficient 
to account for the generation of charged matter expected from anomaly cancellation.

For the other incomplete $E_8$ singularities and complete singularities, 
the singularities remaining after the two small resolutions are not the good kind 
of singularities that can be resolved by additional small resolutions.

\section{Esole-Yau resolution revisited : Proper transform/constraint duality}
\label{duality}
\subsection{Esole-Yau resolution revisited}
In the previous section, we saw that by adopting a different small resolution than 
the one  discussed in the original Esole-Yau paper, 
we can obtain a smooth model up to a certain limit on the number of generic 
codimension-two singularities that gather there, 
although the number of exceptional curves often falls short of the number expected from 
anomaly cancellation. We also found that when the number of generic codimension-two singularities 
that gather there exceeds this limit, a type of singularity appears that cannot be resolved 
by a small resolution.
%
%
On the other hand, we are led to the conclusion in section \ref{thefirstlook} 
that nothing happens even when $G'$ or the coalescence pattern changes 
if we use one of the Esole-Yau small resolutions to resolve multiply enhanced singularities.
In this section, we consider how, or why, this seemingly puzzling difference arises.
Putting the answer first, when a singularity arises that cannot be resolved by 
a small resolution in the alternative way of resolution, 
the {\em constraint condition} becomes singular in the Esole-Yau case.

Since the proper transform of the equation of the threefold after 
the Esole-Yau small resolution takes the form (\ref{Phizx34}) with a constraint (\ref{constraint}),
it can be regarded as a complete intersection Calabi-Yau (though it is ``complete'' only 
in the generic case where there are no multiply enhanced singularities) 
defined by the two equations 
\beqa
\Phi_{zx\xi\zeta}((V_1:U_1),\xi,(V_2:U_2),\zeta,w) 
&\equiv
&     -  V_1 V_2+  U_1 U_2 \left(\xi U_1 + a_{2,1} (w) + a_{4,3}(w)\zeta U_2  +a_{6,5}(w) (\zeta U_2)^2\right)\n
&=&0,
\label{Phizx34again}\\
\Psi_{EY}((V_1 :U_1),\xi,(V_2:U_2),\zeta,w) &\equiv&
-\xi V_1  - a_{1,0}(w) -a_{3,2}(w) \zeta U_2+\zeta V_2\n&=&0
\eeqa
in the {\em five}-dimensional ambient space with 
coordinates $((V_1:U_1),\xi,(V_2:U_2),\zeta,w)$.
$((0:1),0,(0:1),0,0)$ is the point in question and can only be seen in the $U_1=U_2=1$ patch, 
so we set them so below. So let
\beqa
\Phi_{zx\xi\zeta}(V_1,\xi,V_2,\zeta,w)
&\equiv&\Phi_{zx\xi\zeta}((V_1:1),\xi,(V_2:1),\zeta,w) 
\n
&=&     -  V_1 V_2+  \xi  + a_{2,1} (w) + a_{4,3}(w)\zeta   +a_{6,5}(w) \zeta ^2\n
&=&0,
\label{Phizx34again}\\
\Psi_{EY}(V_1 ,\xi,V_2,\zeta,w) &\equiv&\Psi_{EY}((V_1:1),\xi,(V_2:1),\zeta,w) \n
&=&
-\xi V_1 - a_{1,0}(w) -a_{3,2}(w) \zeta +\zeta V_2\n&=&0.
\label{constraintagain}
\eeqa
The manifold $\Psi_{EY}(V_1,\xi,V_2,\zeta,w)=0$ is singular at 
\beqa
\xi=\zeta=V_1=0,~~~V_2=a_{3,2}(w),~~~a_{1,0}(w)=a_{1,0}'(w)=0,
\label{Psisingularity}
\eeqa
which is codimension-five in the five-dimensional ambient space. 
Therefore, it does not in general exist on the manifold $\Phi_{zx\xi\zeta}=0$, and is therefore harmless.
However,  
if $(V_1,\xi,V_2,\zeta,w)=(0,0,a_{3,2}(w_0),0,w_0)$ that satisfies (\ref{Psisingularity})
also happens to satisfy 
\beqa
a_{2.1}(w_0)=0,
\eeqa
then this singularity of $\Psi_{EY}=0$ is also on $\Phi_{zx\xi\zeta}=0$.

What happens if the constraint $\Psi_{EY}=0$ is singular? If $\Psi_{EY}=0$ is regular 
at some point $P$ on the threefold defined as the intersection 
$\{\Phi_{zx\xi\zeta}=0\}\bigcap \{\Psi_{EY}=0\}$, then 
the manifold $\Psi_{EY}=0$ has a tangent hyperplane at $P$ in the five-dimensional ambient space, 
so the derivative of any particular one of the five coordinate variables with respect to the other four 
can be determined by implicit differentiation. 
If, on the other hand, $\Psi_{EY}=0$ is singular 
at $P$, then no such tangent hyperplane can be defined, so $\Psi_{EY}=0$ 
cannot be solved for one of the five coordinates as an implicit function of 
the other four. 
This is exactly what happened in the Esole-Yau resolution 
applied to multiply enhanced singularities considered in this paper.

\subsection{Equivalence of the two models: Proper transform/constraint duality}
In fact, much more can be said. 
From (\ref{Phizxxieta}), we can write the proper transform of the threefold equation 
via our alternative small resolution in the $U_1=V_1'=1$ patch
\beqa
\Phi_{zx\xi\eta}(\eta,\xi,U_3,z_2,w) 
&\equiv& -\xi \eta - a_{1,0}(w)+   z_2(U_3-a_{3,2}(w))  \n
&=&0,
\label{Phizxxieta5D}
\eeqa
and the constraint (\ref{xiconstraint})
\beqa
\Psi(\eta,\xi,U_3,z_2,w) &\equiv&
-\eta U_3+\xi+
a_{2,1}(w)  + a_{4,3}(w)z_2  +a_{6,5}(w) z_2^2\n
&=&0.
\label{xiconstraint5D}
\eeqa

By noticing the fact that $\zeta=z_2$ in the $U_2=1$ patch and 
$\eta=V_1$ in the $V_1'=1$ patch (see (\ref{EYsmallresolution2specific}) and \eqref{eq:u3V1}), 
and comparing (\ref{Phizxxieta5D}), (\ref{xiconstraint5D})
with 
(\ref{Phizx34again}),
(\ref{constraintagain}), we find
\beqa
\Phi_{zx\xi\eta}(\eta,\xi,U_3,z_2,w) &=&\Psi_{EY}(\eta,\xi,U_3,z_2,w),\n
\Psi(\eta,\xi,U_3,z_2,w) &=&\Phi_{zx\xi\zeta}(\eta,\xi,U_3,z_2,w),
\label{eq:duality}
\eeqa
%
%
%
%
where we have made an identification $V_2=U_3$.
In other words, the proper transform of the threefold equation after the 
Esole-Yau small resolution coincides with the constraint equation in the alternative 
small resolution discussed in section \ref{Alternativesmallresolution}, and 
the constraint equation in the Esole-Yau small resolution is equal to the 
proper transform of the threefold equation after the alternative small resolution!

Thus, the two ways of small resolutions are completely equivalent, 
regardless of whether there are multiply enhanced singularities or not; 
both should reach exactly the same conclusions.
In section \ref{thefirstlook}  we encountered the puzzling fact that nothing happened 
in the proper transform, but then we had to resolve the singularities on the 
{\em constraint}, not the proper transform of the threefold equation itself.


In fact, it is easy to see that a change of the center of a small resolution 
(that does not cause a flop) generally results in an interchange of the proper transform 
and the constraint. Let us conclude this section by showing this.
Figure \ref{Fig:Duality} below illustrates this duality.

Suppose that we are given a 
conifold-like binomial equation
\beqa
u_1 u_2&=&v_1 v_2.
\eeqa
If we choose $u_1=v_1=0$ as the center of the small resolution, the change of coordinates is 
\beqa
(u_1,v_1)&=&(\eta U_1, \eta V_1),
\eeqa 
where, as usual, $(U_1:V_1)$ are the projective coordinates and 
$\eta$ is the section of the line bundle for projectivization.
The proper transform is 
\beqa
U_1 u_2&=&V_1 v_2,
\eeqa
which in the $V_1=1$ patch reduces to 
\beqa
U_1u_2&=&v_2.
\label{U1u2=v2}
\eeqa
Since  $\eta$ is $v_1$ itself in this patch, the constraint relating the old variable to the new one is 
\beqa
u_1&=&v_1 U_1.
\label{u1=v1U1}
\eeqa  
If, on the other hand, we take $u_2=v_2=0$ as the center of the small resolution instead, 
we have
\beqa
(u_2,v_2)&=&(\zeta U_2, \zeta V_2),
\eeqa 
where $(U_2:V_2)$ and $\zeta$ are similarly defined.
In this case, the proper transform of the blown-up equation is
given in the $U_2=1$ patch
\beqa
u_1&=&v_1 V_2.
\label{u1=v1V2}
\eeqa
$\zeta$ is equal to $u_2$ this time, where the constraint 
\beqa
v_2&=&u_2 V_2
\label{v2=u2V2}
\eeqa
is 
imposed.
Comparing 
(\ref{U1u2=v2}), (\ref{u1=v1U1}) 
with
(\ref{u1=v1V2}), (\ref{v2=u2V2}),
we see that, with an identification $U_1=V_2$,
 the proper transform (\ref{U1u2=v2}) is equivalent to the constraint 
 (\ref{v2=u2V2}), 
 and the constraint (\ref{u1=v1U1}) is the same as the proper transform (\ref{u1=v1V2}).
The same holds true for the $V_2=1$ and $U_1=1$ patches.

\begin{figure}[htb]
  \begin{center}          
         \includegraphics[clip, width=12.4cm]{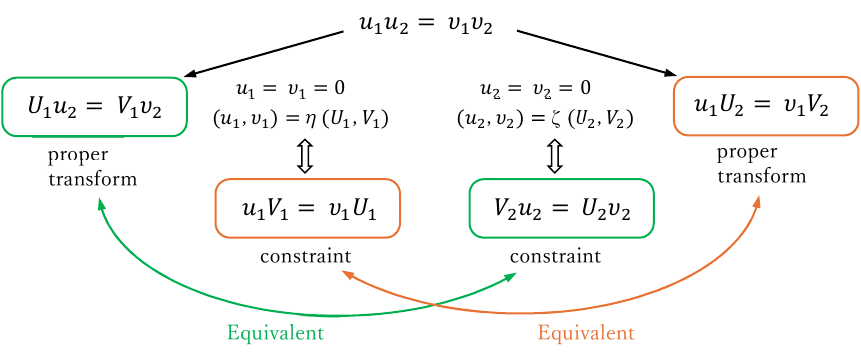}
       \caption{The proper transform/constraint duality for the conifold singularity.}
    \label{Fig:Duality}
  \end{center}
\end{figure}

\section{Summary and conclusions}
In this paper, we investigated the geometrical structure of multiply enhanced codimension-two singularities in the $SU(5)$ model of six-dimensional F-theory, where the rank of the singularity increases by two or more.
Multiply enhanced singularities do not exist generically, but arise at special points in the moduli space where several ordinary codimension-two singularities gather and overlap.
There are various patterns in how such singularities gather, and the charged matter that should be generated there can be predicted based on the anomaly cancellation conditions.
In this paper, we have performed blow-up processes to verify whether the exceptional curves that can explain the predicted generation of charged matter emerge through the resolution of the multiply enhanced singularities for each case where the singularity is enhanced from $SU(5)$ to $E_6$, $E_7$, and $E_8$.

We first applied one of the six small resolutions developed by Esole-Yau to the multiply enhanced singularities.
However, it was observed that the proper transform of the threefold equation obtained in this way does not reflect changes in the singularity or how the generic codimension-two singularities gather there.
Therefore, we resolved the multiply enhanced singularities by changing the center of the last small resolution of Esole-Yau.
This change of the center would have resulted in the insertion of an equivalent $\PP^1$ (different from the flop) 
as a small resolution of a conifold.
However, when we actually went through the resolution procedure, we obtained a threefold proper transform that was different from the result obtained by the first way, 
and this depended precisely on the multiply enhanced singularities and the way the generic codimension-two singularities overlap.


The results are:
\begin{itemize}
\item{
In the case of $G'=E_6$, after the two small resolutions, the threefold equation becomes regular for the incomplete singularity, and a conifold singularity remains for the complete singularity.
In 
both cases, the resolutions yield sets of exceptional curves consistent with the spectrum expected from anomaly cancellation.}
\item{
In the case of $G' = E_7$, similarly after the two small resolutions, the threefold equation becomes regular in the case of the incomplete $1$ singularity, and a set of exceptional curves consistent with anomaly cancellation is obtained.
We saw that in the other incomplete cases $2$,\,$3$ and $4$, a conifold singularity remains.
It is probably enough to complete the set of exceptional curves needed to cancel the anomalies 
for the incomplete $2$ case, but not enough for the incomplete $3$ and $4$ cases.
In the case of the complete singularity, there arises a type of singularity that is neither a conifold nor a generalized conifold singularity.
If this is resolved, the canonical class will not be preserved, and so if this transition actually proceeds, supersymmetry will be broken.
}

\item{
In the case of $G'=E_8$, it was found that after the two small resolutions, even the incomplete $1$ singularity only yields exceptional curves that are insufficient to cancel the anomaly.
For the other incomplete and complete singularities, it was shown that either a conifold singularity appears and can be resolved but is insufficient to cancel the anomaly, or a singularity reappears that cannot be resolved by small resolutions.
}
\end{itemize}

Finally, we revisited why the first Esole-Yau small resolution did not yield these results.
As a result, it turned out that the change of the center that brings about the difference between the two ways of small resolutions actually leads to an interchange of the proper transform and the constraint condition, and under this interchange the two ways of small  resolutions are completely equivalent.
Therefore, even in the Esole-Yau small resolution, the same conclusion could have been reached if the constraint equation rather than the proper transform had been resolved by small resolutions.

As we noted in the text, the fact that the number of exceptional curves fall short 
for anomaly cancellation does not mean that the anomaly 
cancellation breaks down. 
In the present case, several exceptional curves 
overlap to form an identical single curve;
it remains to be seen in the future what kind of hypermultiplets arise when, 
as we have seen in this paper, several exceptional curves overlap to form 
an identical single curve, since the anomalies should cancel out anyway.

One of the original motivations of this research is to examine whether 
our previous proposal \cite{MTanomaly} to realize the Kugo-Yanagida 
$E_7/(SU(5)\times SU(3) \times U(1))$ 
K\"ahler coset using localized massless matter in F-theory is really possible.
The results of this paper give a negative answer to that question, 
at least in six dimensions.

In this paper, we have investigated the multiple singularity enhancement in F-theory 
exclusively for compactifications with unbroken gauge symmetry SU(5) in six dimensions, 
but it is expected that similar results will be obtained when the unbroken gauge symmetry 
is different from SU(5).

On the other hand, when applying the discussion to four dimensions, the results of the 
six-dimensional F-theory compactification obtained here do not immediately apply. 
Of course, in four dimensions, rank-two enhancement at codimension-three singularities 
inevitably occurs, and it has already been observed that the exceptional curves there do 
not yield ``honest'' Dynkin diagram of the apparent Kodaira fiber type dictated by the 
various zero orders of the relevant sections there \cite{Yukawas}. However, four-dimensional 
F-theories are essentially different from six-dimensional ones in the following two points.

The first difference is that in four dimensions the chiral matter spectrum is not only determined by 
geometric peculiarities, but also by the so-called G-flux \cite{Curio,AndreasCurio,DonagiWijnholt2}. 
The second difference is that in four dimensions the anomaly cancellation conditions are not as strict as 
in six dimensions, and it is possible for a chirality changing transition to occur while maintaining the 
anomaly-free condition \cite{Curio,chiralitychangingtransition1,chiralitychangingtransition2}. 
Therefore, the analysis of multiple 
enhancement in four dimensions needs to be carried out separately. We hope to come back to this 
issue in the near future.

\section*{Acknowledgement}
We thank Yuta Hamada for useful discussions.
The work of S.M. was supported by JSPS KAKENHI Grant Number JP23K03401
and the work of T.T. was supported by JSPS KAKENHI Grant Number JP22K03327.

\newpage

\section*{Appendix A}
\begin{table}[htbp]
\centering
\begin{tabular}{|ccccc|c|c|}
\hline
$\mbox{o}_w(a_{1,0})$  &  $\mbox{o}_w(a_{2,1})$ &
$\mbox{o}_w(a_{3,2})$ &  $\mbox{o}_w(a_{4,3})$ &
$\mbox{o}_w(a_{6,5})$ &  $\mbox{o}_w(P_{8,5})$& \mbox{name} \\  \hline   \vspace{-0.1cm}   
 $1$ & $1$ & $1$ & $1$ &$0$& $2$ & incomplete  1 \\                         \vspace{-0.1cm}
 $2$ & $1$ & $1$ & $1$ &$0$& $3$ & incomplete  2 \\                         \vspace{-0.1cm}
 $2$ & $1$ & $2$ & $1$ & $0$ & $4$ & incomplete 3 \\                        \vspace{-0.1cm}
 $2$ & $2$ & $1$ & $1$ & $0$ & $4$ & incomplete 4  \\                       \vspace{-0.1cm}
 $3$ & $1$ & $1$ & $1$ & $0$ & $3$ & incomplete  5 \\                       \vspace{-0.1cm} 
 $3$ & $1$ & $2$ & $1$ & $0$ & $5$ & incomplete  6 \\                       \vspace{-0.1cm}
 $3$ & $1$ & $3$ & $1$ & $0$ &  $6$  & incomplete 7\\                       \vspace{-0.1cm}
 $3$ & $2$ & $1$ & $1$ & $0$ & $4$ & incomplete 8\\                         \vspace{-0.1cm}
 $3$ & $2$ & $2$ & $1$ & $0$ &  $6$  & incomplete 9\\                       \vspace{-0.1cm}
 $3$ & $3$ & $1$ & $1$ & $0$ & $5$ & incomplete 10\\                       \vspace{-0.1cm}
 $3$ & $4$ & $1$ & $2$ & $0$ &  $6$  & incomplete 11\\                     \vspace{-0.1cm} 
 $4$ & $1$ & $1$ & $1$ & $0$ & $3$ & incomplete 12\\                       \vspace{-0.1cm}
 $4$ & $1$ & $2$ & $1$ & $0$ & $5$ & incomplete 13\\                       \vspace{-0.1cm}
 $4$ & $1$ & $3$ & $1$ & $0$ &  $7$ & incomplete 14\\                      \vspace{-0.1cm}
 $4$ & $1$ & $4$ & $1$ & $0$ & $8$ & incomplete 15\\                       \vspace{-0.1cm}
 $4$ & $2$ & $1$ & $1$ & $0$ & $4$ & incomplete 16\\                       \vspace{-0.1cm}
 $4$ & $2$ & $2$ & $1$ & $0$ &  $6$  & incomplete 17  \\                   \vspace{-0.1cm}  
 $4$ & $2$ & $3$ & $1$ & $0$ & $8$ & incomplete 18  \\                     \vspace{-0.1cm}
 $4$ & $3$ & $1$ & $1$ & $0$ & $5$ & incomplete 19  \\                     \vspace{-0.1cm}
 $4$ & $3$ & $2$ & $1$ & $0$ &  $7$ & incomplete  20 \\                    \vspace{-0.1cm}
 $4$ & $4$ & $1$ & $1$ & $0$ &  $6$  & incomplete  21 \\                   \vspace{-0.1cm}
 $4$ & $4$ & $2$ & $2$ & $0$ & $8$ & incomplete 22  \\                     \vspace{-0.1cm}
 $4$ & $5$ & $1$ & $2$ & $0$ &  $7$ & incomplete 23 \\                     \vspace{-0.1cm}
 $4$ &  $6$  & $1$ & $3$ & $0$ & $8$ & incomplete 24 \\                    \vspace{-0.1cm}
 $5$ & $1$ & $1$ & $1$ & $0$ & $3$ & incomplete 25 \\                      \vspace{-0.1cm}
 $5$ & $1$ & $2$ & $1$ & $0$ & $5$ & incomplete 26 \\                      \vspace{-0.1cm}
 $5$ & $1$ & $3$ & $1$ & $0$ &  $7$ & incomplete 27 \\                     \vspace{-0.1cm}
 $5$ & $1$ & $4$ & $1$ & $0$ & $9$ & incomplete  28\\                      \vspace{-0.1cm}
 $5$ & $1$ & $5$ & $1$ & $0$ & $10$ & complete 1  \\                       \vspace{-0.1cm}         
 $5$ & $2$ & $1$ & $1$ & $0$ & $4$ & incomplete  29\\                      \vspace{-0.1cm}
 $5$ & $2$ & $2$ & $1$ & $0$ &  $6$  & incomplete 30 \\                    \vspace{-0.1cm}
 $5$ & $2$ & $3$ & $1$ & $0$ & $8$ & incomplete 31 \\                      \vspace{-0.1cm}
 $5$ & $2$ & $4$ & $1$ & $0$ & $10$ & complete 2\\                         \vspace{-0.1cm}       
 $5$ & $3$ & $1$ & $1$ & $0$ & $5$ & incomplete  32\\                      \vspace{-0.1cm}
 $5$ & $3$ & $2$ & $1$ & $0$ &  $7$ & incomplete  33\\                     \vspace{-0.1cm}
 $5$ & $3$ & $3$ & $1$ & $0$ & $9$ & incomplete  34\\                      \vspace{-0.1cm}
 $5$ & $4$ & $1$ & $1$ & $0$ &  $6$  & incomplete  35\\                    \vspace{-0.1cm}
 $5$ & $4$ & $2$ & $1$ & $0$ & $8$ & incomplete  36\\                      \vspace{-0.1cm}
 $5$ & $4$ & $3$ & $2$ & $0$ & $10$ & complete 3 \\                        \vspace{-0.1cm}      
 $5$ & $5$ & $1$ & $1$ & $0$ &  $7$ & incomplete  37\\                     \vspace{-0.1cm}
 $5$ & $5$ & $2$ & $2$ & $0$ & $9$ & incomplete  38\\                      \vspace{-0.1cm}
 $5$ &  $6$  & $1$ & $2$ & $0$ & $8$ & incomplete  39\\                    \vspace{-0.1cm}
 $5$ &  $6$  & $2$ & $3$ & $0$ & $10$ & complete 4\\                       \vspace{-0.1cm}      
 $5$ &  $7$ & $1$ & $3$ & $0$ & $9$ & incomplete  40\\                     \vspace{-0.0cm}
 $5$ & $8$ & $1$ & $4$ & $0$ & $10$ & complete  5\\                 
\hline
\end{tabular}
\caption{$E_8$ patterns. The anomaly-free spectrum for each case is given by 
$\mbox{o}_w(a_{1,0}) \cdot {\bf 10} \oplus \mbox{o}_w(P_{8,5})\cdot {\bf 5}$ (see \eqref{eq:anomalyfreeHc}).}
\label{E8pattern}
\end{table}
%


\end{document}